\documentclass[preprint,5p,times,twocolumn]{elsarticle}
\makeatletter
\def\ps@pprintTitle{%
 \let\@oddhead\@empty
 \let\@evenhead\@empty%
 \let\@evenfoot\@oddfoot}
\makeatother
\usepackage[utf8]{inputenc}
\usepackage{soul}
\usepackage{amsmath}
\usepackage[english]{babel}
\usepackage{siunitx}
\usepackage{wrapfig}
\usepackage[switch,modulo]{lineno}
\usepackage{caption}
\usepackage{xparse}
\ExplSyntaxOn
\NewDocumentCommand{\longdash}{ O{2} }
 {
  --\prg_replicate:nn { #1 - 1 } { \negthinspace -- }
 }
\ExplSyntaxOff
\usepackage{subcaption}
\usepackage{tikz}
\usetikzlibrary{decorations.pathreplacing,calligraphy}
\usepackage{geometry}                		
\usepackage{graphicx}				

\graphicspath{ {figures/}  } 

\usepackage{multirow}
\usepackage{epstopdf}                 
\DeclareGraphicsExtensions{.png,.jpg}  

\usepackage{booktabs}
                  
\usepackage[backref=page,         
            pagebackref=false,    
            hyperindex=true,      
            colorlinks=true,      %
            citecolor=blue,       %
            filecolor=black,      %
            linkcolor=blue,       %
            urlcolor=blue,        %
            breaklinks=true,      %
            bookmarks=true,       
            bookmarksopen=false,  
            pdftitle   = {},
            pdfauthor  = {},
            pdfsubject = {}
]{hyperref} 

\usepackage{amssymb}
\usepackage{lipsum}                   
\usepackage{textcomp}
\usepackage{booktabs}
\usepackage{xcolor}

\definecolor{goold}{RGB}{255,200,0}

\usepackage{natbib}
\setcitestyle{square,numbers}
\begin{document}
\title{Application of recoil-imaging time projection chambers to directional neutron background measurements in the SuperKEKB accelerator tunnel}
\author[add1]{J. Schueler\corref{cor1}}
\ead{jschuel@hawaii.edu}
\cortext[cor1]{Corresponding author}
\author[add1]{S.~E.~Vahsen}
\address[add1]{Department of Physics and Astronomy, University of Hawaii, Honolulu, Hawaii 96822, USA}
\author[add1]{P.~M.~Lewis\fnref{fn1}}
\fntext[fn1]{Now at University of Bonn, 53012 Bonn, Germany}
\author[add1]{M.~T.~Hedges\fnref{fn2}}
\fntext[fn2]{Now at Purdue University, West Lafayette, Indiana 47907, USA}
\author[add5,add6]{D. Liventsev}
\address[add5]{Department of Physics and Astronomy, Wayne State University, Detroit, Michigan 48201, USA}
\author[add9]{F. Meier}
\author[add6]{H. Nakayama}
\address[add6]{High Energy Accelerator Research Organization (KEK), Tsukuba, 305-0801 Japan}
\author[add1]{A. Natochii}
\author[add1]{T.~N.~Thorpe\fnref{fn5}}
\fntext[fn5]{Now at University of California Los Angeles, Los Angeles, California, 90095, USA}
\address[add9]{Duke University, Durham, North Carolina 27708, USA}
\date{\today}

\begin{abstract}
Gaseous time projection chambers (TPCs) with high readout segmentation are capable of reconstructing detailed 3D ionization distributions of nuclear recoils resulting from elastic neutron scattering. Using a system of six compact TPCs with pixel ASIC readout, filled with a 70:30 mixture of $\text{He}$:$\text{CO}_2$ gas, we analyze the first directional measurements of beam-induced neutron backgrounds in the tunnel regions surrounding the Belle II detector at the SuperKEKB $e^+e^-$ collider. With the use of 3D recoil tracking, we show that these TPCs are capable of maintaining nearly $100\%$ nuclear recoil purity to reconstructed ionization energies ($E_\text{reco}$) as low as $\SI{5}{keV_{ee}}$. Using a large sample of Monte-Carlo (MC)-simulated $^4\text{He}$, $^{12}\text{C}$, and $^{16}\text{O}$ recoil tracks, we find consistency between predicted and measured recoil energy spectra in five of the six TPCs, providing useful validation of the neutron production mechanisms modeled in simulation. Restricting this sample to $^4\text{He}$ recoil tracks with $E_\text{reco}>\SI{40}{keV_{ee}}$, we further demonstrate axial angular resolutions within $8^{\circ}$ and we introduce a procedure that under suitable conditions, correctly assigns the vector direction to $91\%$ of these simulated $^4\text{He}$ recoils. Applying this procedure to assign vector directions to measured $^4\text{He}$ recoil tracks, we observe consistency between the angular distributions of observed and simulated recoils, providing first experimental evidence of localized neutron ``hotspots" in the accelerator tunnel. Observed rates of nuclear recoils in these TPCs suggest that simulation overestimates the  neutron flux from these hotspots. Despite this, we estimate these hotspots to produce the majority of neutron backgrounds in the accelerator tunnel at SuperKEKB's target luminosity of $\SI{6.3e35}{cm}^{-2}\text{s}^{-1}$, making them important regions to continue to monitor.
\end{abstract}

\maketitle

\section{Introduction}
\label{sec:intro}
The SuperKEKB accelerator \cite{ohnishi} located at the KEK laboratory in Tsukuba, Japan, is a high-luminosity asymmetric-energy circular collider that has been providing $e^+e^-$ collisions for the Belle II $B$-factory experiment \cite{belle} since March 2018, and has held the world record instantaneous luminosity since June 2020. As of December 2021, SuperKEKB has reached an instantaneous luminosity of $\SI{3.8e34}{cm}^{-2}\text{s}^{-1}$  and seeks to ultimately achieve a luminosity of $\SI{6.3e35}{cm}^{-2}\text{s}^{-1}$ \cite{natochiisnowmass}. This ambitious luminosity will be achieved both by roughly doubling the currents of the $e^+$ and $e^-$ beams in the Low Energy Ring (LER) and High Energy Ring (HER), respectively, and by shrinking the vertical beam sizes at the interaction point (IP) by a factor of roughly 20 over SuperKEKB's predecessor, KEKB \cite{kekb}. The reduction of vertical beam sizes is made possible by employing a ``Nano-beam scheme" proposed by P. Raimondi \cite{nanobeam}. Achieving such a high luminosity, however, comes at the cost of elevated beam-induced backgrounds, making it important to understand and properly mitigate each background source to ensure the longevity and stable operation of the Belle II detector. 

Neutron backgrounds are particularly difficult to deal with at high luminosity colliders and were responsible for dead time in the $K_L$-Muon detector (KLM) of Belle, the predecessor to Belle II \cite{klm1,klm2}. Since neutrons are electrically neutral, they lose energy primarily from elastic scattering off of atomic nuclei, making them highly penetrating if not properly shielded by low-$Z$ materials. To prevent a similar scenario in Belle II, polyethylene shielding was installed on the outer KLM endcaps, and the RPC layers on both the outer KLM endcaps and two innermost barrel layers were replaced with scintillator-based detectors tolerant of increased background hit rates \cite{klm3}. Given the expectation of higher background rates at SuperKEKB over KEKB, and that these rate estimates have substantial uncertainties \cite{natochiisnowmass,natochii}, neutrons could pose a risk not only to the KLM, but also to other Belle II detector systems where both fast and thermal neutrons can lead to single event upsets \cite{TOP, SEU}. For these reasons, it is important to understand neutron production at SuperKEKB in order to best assess neutron background remediation measures.

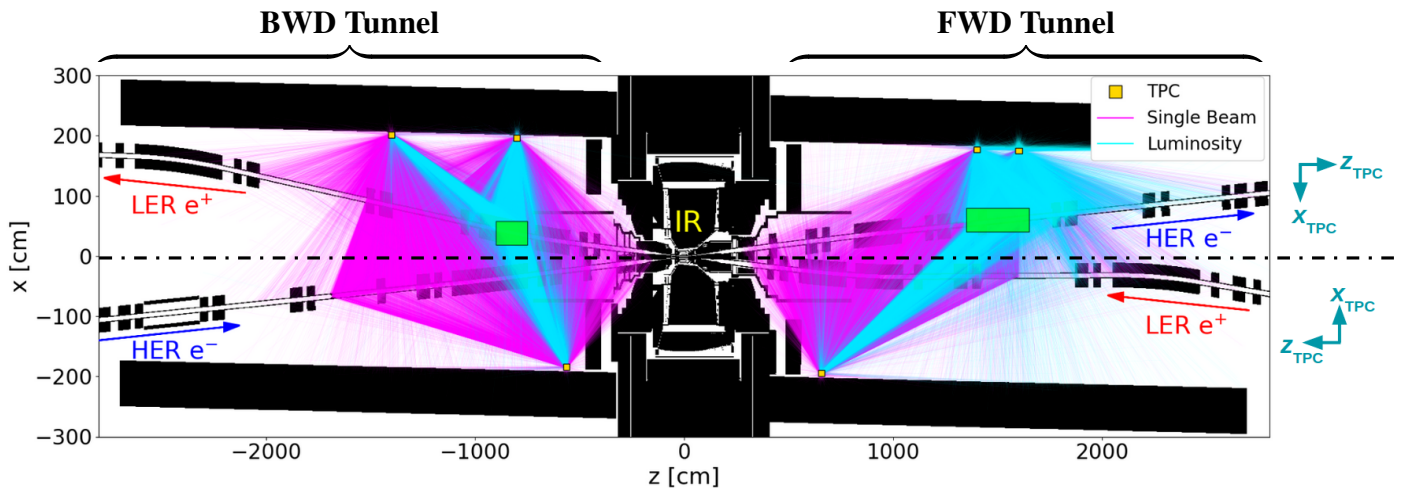
\begin{figure*}[htbp]
\begin{center}
\begin{tikzpicture}[ultra thick, overlay]
\draw [pen colour={black},
    decorate,
    decoration = {calligraphic brace,
        raise=3pt,
        amplitude=8pt},xshift=-3cm,yshift=0.6cm] (-5.0,-0.8) -- (1.7,-0.8) node [black,midway,yshift = 0.7cm]
{\footnotesize \textcolor{black}{\large \textbf{BWD Tunnel}}};

\draw [pen colour={black},
    decorate, 
    decoration = {calligraphic brace,
        raise=3pt,
        amplitude=8pt},xshift=-3cm,yshift=0.6cm] (4.1,-0.8) -- (10.4,-0.8) node [black,midway,yshift = 0.7cm] {\footnotesize \textcolor{black}{\large \textbf{FWD Tunnel}}};
\end{tikzpicture}

\includegraphics[width=\linewidth]{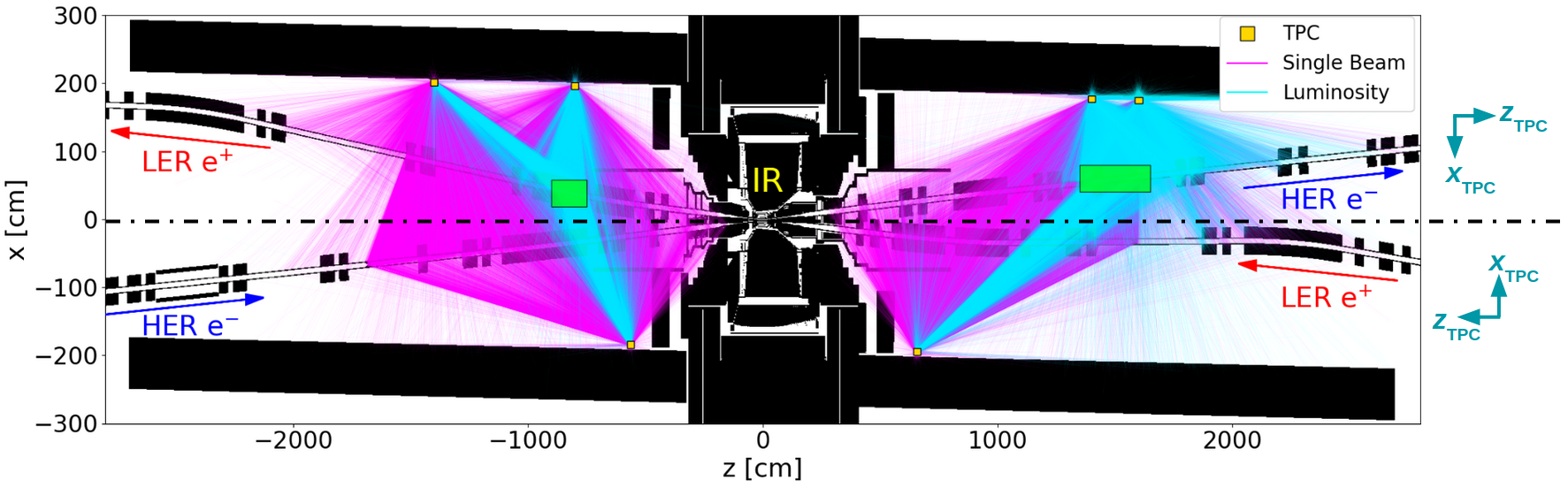}
\caption{(color online) Geant4 material scan of the Belle II detector and the SuperKEKB accelerator rings with each of the six Phase 3 TPCs shown to scale as yellow boxes. The coordinate axes on the far right show local TPC coordinate systems in reference to Belle II coordinates. The magenta and cyan traces represent simulated background neutrons passing through the sensitive volume of a TPC, produced from single beams and collisions (luminosity), respectively. Each trace originates at the production point of the simulated neutron and terminates at the TPC vessel. The majority of luminosity-induced simulated neutrons come from two highly localized regions (``hotspots") shown in green.}
\label{fig:phase3}
\end{center}
\end{figure*}
Neutron backgrounds originate from showers that result from off-orbit beam particles or photons interacting with the atoms in the walls of the beam pipe. Secondaries from these showers excite particular nuclei via the giant dipole resonance \cite{giantres1, giantres2}, producing neutrons. Neutron production can arise both from circulating single beams and from collisions. Beam-gas scattering (Coulomb and Bremsstrahlung) and the Touschek effect are the principal single beam-induced mechanisms leading to neutron-producing showers. Through-going photons, emitted from radiative Bhabha (RBB) scattering events from colliding beams at the IP, travel along the straight section of beam pipes, ultimately colliding with the beam pipe walls in the region where the beam pipes start to curve. These RBB photon collisions are highly localized and produce copious amounts of neutrons leading to what we refer to as \textit{radiative Bhabha hotspots} (green boxes in \def\figureautorefname{Fig.}\autoref{fig:phase3}). Each mechanism leading to neutron production is difficult to simulate, making it important to directly measure neutron backgrounds and use these measurements to both improve our current understanding of neutron backgrounds, and also improve our ability to forecast neutron backgrounds in future scenarios.

With these goals in mind, we have deployed systems of recoil-imaging time projection chambers (TPCs) \cite{nygren} throughout all three beam commissioning Phases of SuperKEKB. These TPCs are capable of producing 3D directional images of ionized gas nuclei resulting from fast neutron scattering. The distinct TPC systems employed during first and second Phases of beam commissioning (known as Phase 1 and Phase 2, respectively) were concerned with measuring fast neutron backgrounds produced near the IP. Of the principal neutron production regions, the tunnel regions surrounding either side of Belle II were not instrumented in Phases 1 or 2, making them the regions of interest for the present Phase 3 TPC system, which has been in operation since March 2019. In this work, we analyze fast neutron background measurements recorded by the Phase 3 TPC system which is comprised of six TPCs: three in the \textit{backward} tunnel (henceforth called BWD, which corresponds to $z_\text{BELLE} < \SI{-4}{m}$) and three in the \textit{forward} tunnel (henceforth called FWD, which corresponds to $z_\text{BELLE} > \SI{4}{m}$; see \autoref{fig:phase3}). In particular, we make direct comparisons between measured and simulated rates, energies, and angular distributions of nuclear recoils in these TPCs. Together, these comparisons provide insight toward the effectiveness of our current modeling of neutron production at SuperKEKB and allow us to assess the neutron background remediation measures (i.e. shielding, beam optics adjustments) needed to ensure stable operation of Belle II at higher luminosity conditions. All reported measurements were recorded during two dedicated background study days conducted on May 9th, 2020 and June 16th, 2021, which we will refer to as Study A and Study B, respectively. Additional descriptions and analyses of background sources at SuperKEKB can be found in Refs. \cite{natochii, hedges, Gabriel, lewis, skb1, Cuesta, deJong, liptak}.

\begin{figure}[htbp]
\begin{center}
\includegraphics[width=0.37\textwidth]{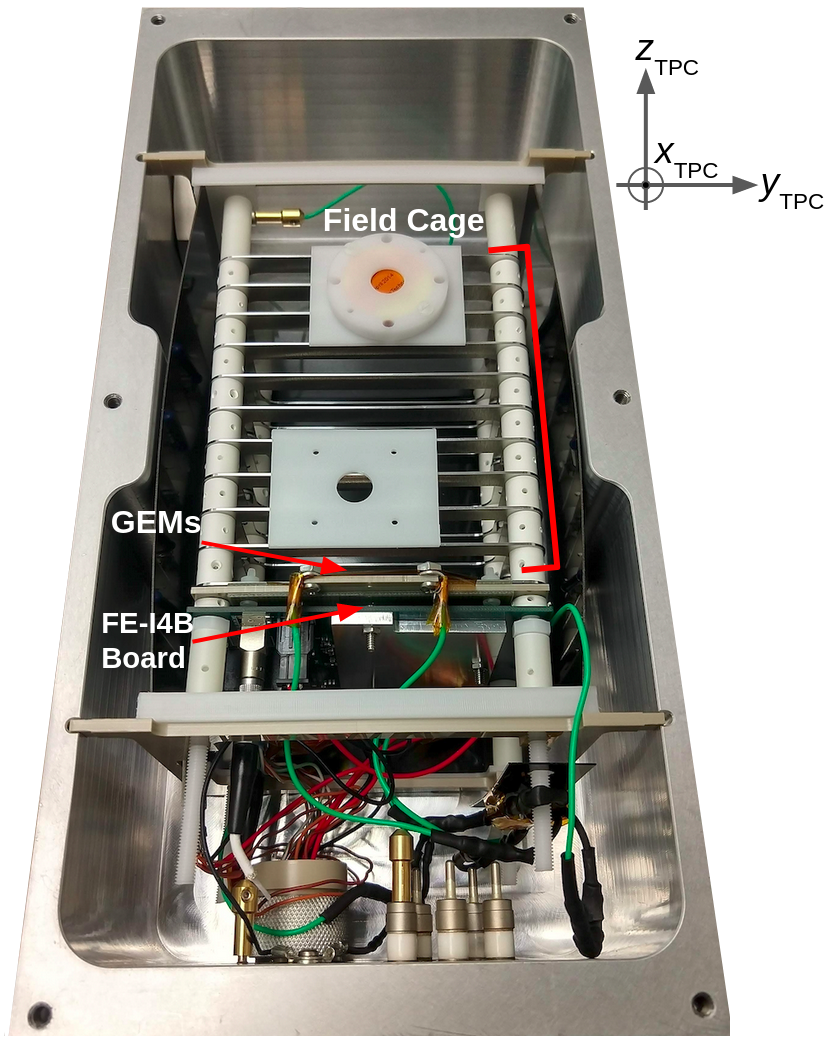}
\caption{(color online) The inside of a BEAST TPC. The key components in the detector are the field cage, the double gas electron multiplier (GEM) layer, and the ATLAS FE-I4B pixel chip. The white disk mounted on top of the field cage contains a $^{210}\text{Po}$ alpha emitting source that is used to calibrate the energy scale of the TPC (\def\subsectionautorefname{Section}\autoref{subsec:gain}). The coordinate axes shown define the TPC coordinate system. The origin is placed on the edge of the readout plane of the pixel chip, with positive $z_\text{TPC}$ pointing upward toward the field cage.}
\label{fig:tpc}
\end{center}
\end{figure}

\section{The TPC system}
\label{sec:TPCs}

The Phase 3 directional fast neutron detection system consists of six BEAST TPCs \cite{Jaegle}, and their high voltage (HV), low voltage (LV), gas, and data acquisition (DAQ) systems. \autoref{fig:phase3} shows the locations of each of the TPCs drawn to scale as yellow boxes. The location of each TPC was chosen with the hope that the TPC would be sensitive to measuring background neutrons generated from the predicted RBB hotspots.

\begin{figure}[htbp]
\begin{center}
\includegraphics[width=0.5\textwidth]{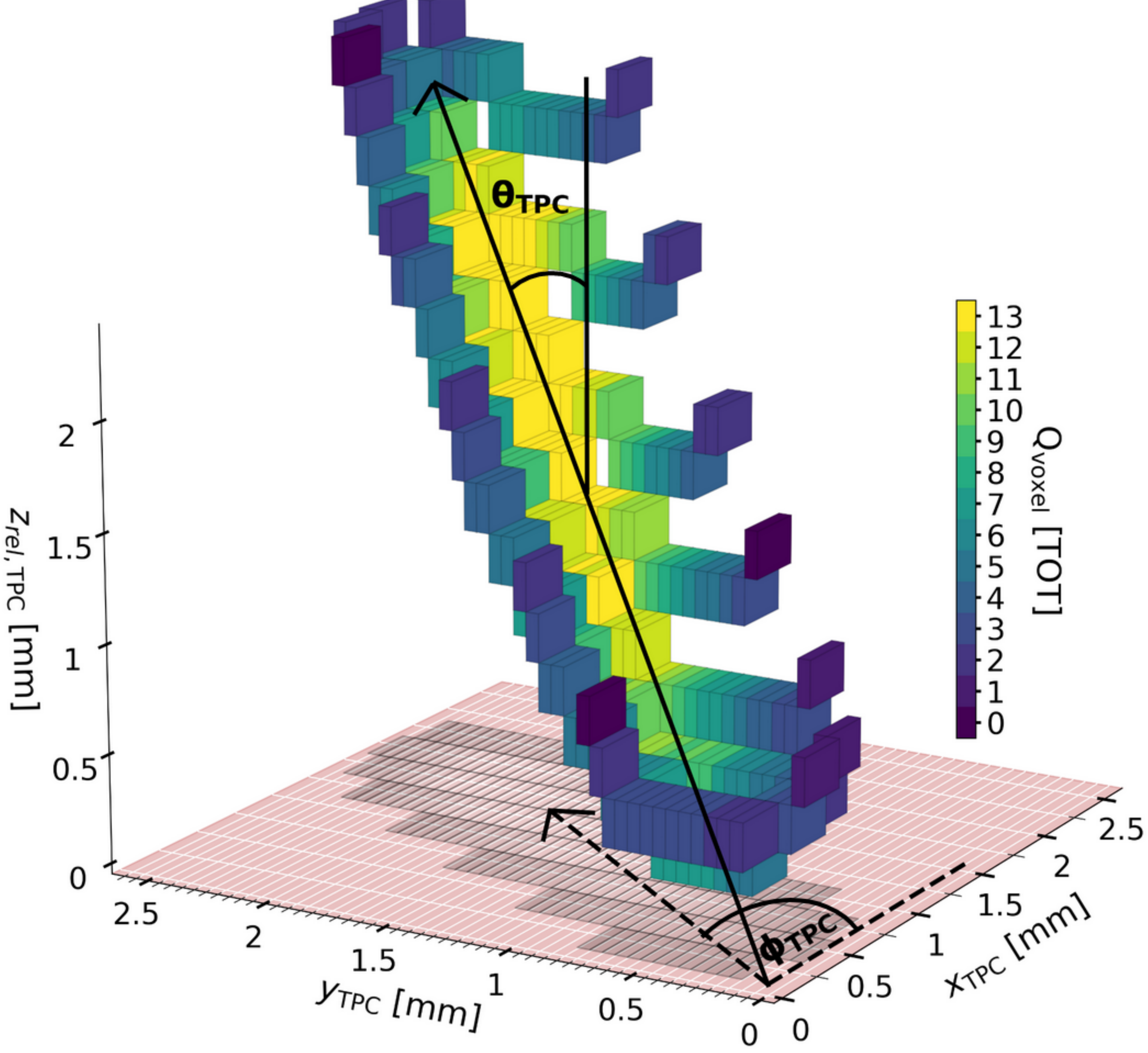}
\caption{(color online) Visualization of a $\SI{210}{keV_{ee}}$ nuclear recoil track reconstructed into 3D voxels. The color scale shows the charge present in each voxel in units of time over threshold (TOT) (\def\sectionautorefname{Section}\autoref{sec:calibration}). $\theta_\text{TPC}$ is the zenith angle defined between $z_\text{TPC}$ and the vector direction of principal axis of the track, and $\phi_\text{TPC}$ is the azimuthal angle defined between $x_\text{TPC}$ and the projection of the track's principal axis onto the readout plane of the chip. The white grid marks illustrate the $\SI{250}{\um}\times\SI{50}{\um}$ pixel dimensions.}
\label{fig:belle2_coord}
\end{center}
\end{figure}

Each TPC is a $15\times 10\times\SI{31}{cm}^3$ vessel with a $2\times 1.68\times\SI{ 10}{cm}^3$ fiducial volume. The vessels are filled with a 70:30 mixture of $\text{He}$:$\text{CO}_2$ gas, which serves as the target gas with which neutrons interact. When a fast neutron enters a TPC, it may scatter off of a $^4\text{He}$, $^{12}\text{C}$, or $^{16}\text{O}$ nucleus, causing the nucleus to recoil. The recoiling nucleus will then collide with other gas atoms in the vessel until it eventually stops. Along the way, the recoiling nucleus strips electrons off of the gas atoms, forming an ionization trail. A uniform electric field along $z_\text{TPC}$ (see \def\figureautorefname{Figs.}\autoref{fig:phase3}, \ref{fig:tpc}, and \ref{fig:belle2_coord}\def\figureautorefname{Fig.} for TPC and Belle II coordinate system definitions) is provided inside an aluminum field cage, causing the ionized nuclei to drift upward toward the cathode, while the electrons in the ionization trail drift and diffuse--both transversely and longitudinally--against the field toward a double gas electron multiplier (GEM) layer \cite{Sauli} where the charge is avalanche-multiplied. This avalanche-multiplied charge is then read out onto the two dimensional readout plane of an ATLAS FE-I4B pixel chip \cite{ATLAS1, ATLAS2}. The constant drift speed (on average) of the ionization charge through the field cage volume allows for the construction of a relative $z$ coordinate, providing a 3D reconstruction of the ionization distribution created by the recoiling gas nucleus (\autoref{fig:belle2_coord}).

During production, the BEAST TPCs were demonstrated to be capable of effective gains up to 50,000, enabling detection of single electrons. However here, we operate the detectors at comparatively low effective gains of $\mathcal{O}(1000)$ to avoid saturation due to the limited dynamic range of the pixel chip when detecting highly ionizing nuclear recoils. This choice also delays the onset of gas detector aging and lowers the risk of accidental sparking which might damage the sensitive pixel electronics. We have now operated these detectors continuously for nearly four years, at times in extreme background conditions,  without any indication of electronics damage or aging effects, so this low-gain operating strategy appears to be successful.
\begin{figure*}[htbp]
\begin{center}
\includegraphics[width=\textwidth]{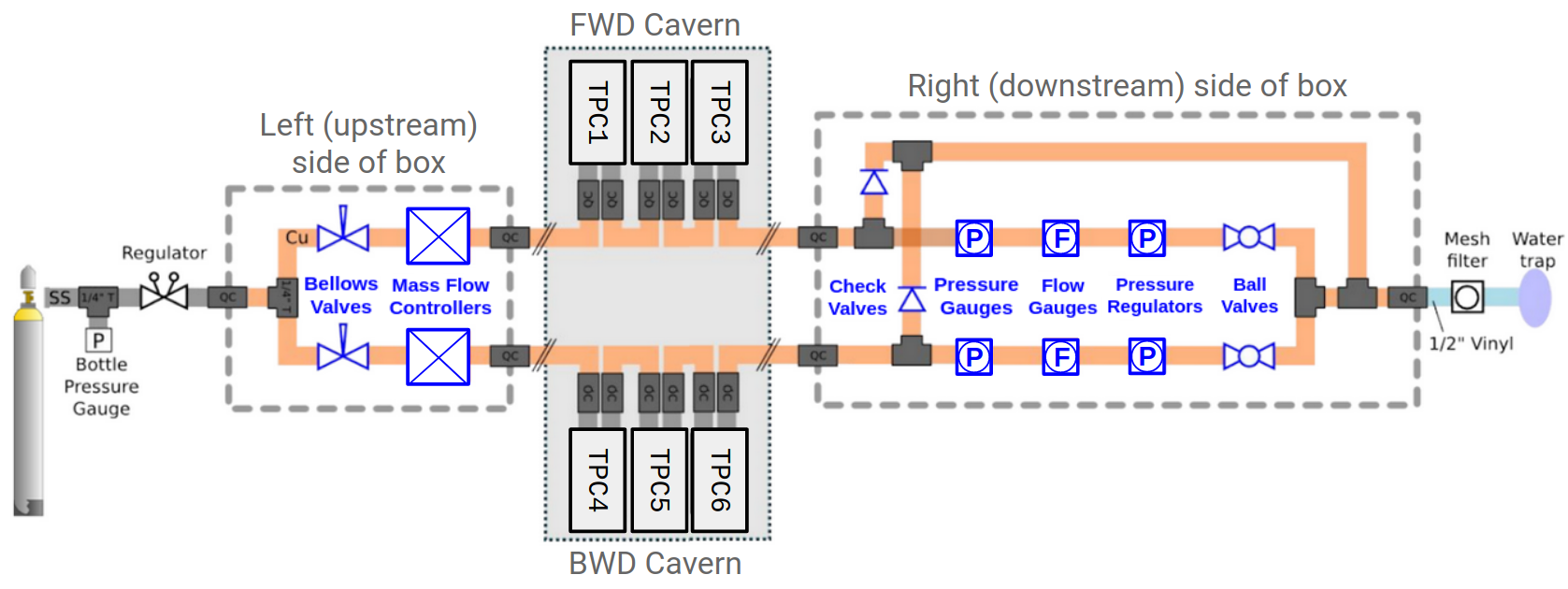}
\caption{(color online) Schematic of the Phase 3 gas system [not to scale]. The components shown within the gray dashed lines reside within a single box that serves as the central gas control hub for the Phase 3 TPC system. This \textit{gas box} was also used to control the gas system in Phase 2.}
\label{fig:gas}
\end{center}
\end{figure*}

\subsection{Gas system}
Maintaining adequate gas purity in each TPC is essential for maintaining a stable effective gain in the GEMs, a necessary condition for optimally measuring and identifying nuclear recoils. During Phase 1 operation, we flowed gas in parallel to four detectors. We auto-regulated the pressure in each detector separately, controlled only the total flow to all detectors, and mechanically adjusted flow impedance with valves to balance parallel flows. This approach was highly unstable, with large flow oscillations and even occasional reverse flow in some of the parallel branches. Gas purity and gain stability, however, were better than expected: although there was a brief initial period of outgassing where gain would slowly rise, after several months of detector operation with gas flow, the avalanche gain would remain stable for weeks even at minimal or at no flow. The Phase 2 and Phase 3 gas systems (\autoref{fig:gas}) were re-designed based on this experience. Two parallel flow branches now have separately controlled flows to avoid oscillations, with the serial flow rate set between 15 and \SI{20}{sccm} through the three TPCs present in each branch. A 70:30 mixture of He:CO$_2$ with a minimum purity of 99.999$\%$ has been used throughout all three Phases of beam commissioning.

\subsection{DAQ and low voltage system}
To communicate between the TPCs and the DAQ system (henceforth referred to as the DAQ box), we construct a custom LV cable assembly for each TPC. These cables route power and low voltage differential signal (LVDS) communications between the FE-I4B chip and the DAQ box. Due to cable length restrictions with LVDS communication, two identical but independent DAQ boxes are employed: one for the three TPCs in the BWD tunnel and one for the three TPCs in the FWD tunnel. A custom Multi-chip Module Card (MMC3) baseboard with an attached Xilinx Kintex-7 FPGA is housed in each DAQ box and serves as the conduit for communication between a PC and the TPC.

\subsection{High voltage system}
Each TPC contains two high voltage (HV) inputs: one SHV input used to power the gas electron multipliers (GEMs) via an internal voltage divider circuit \cite{Jaegle}, and one UHV input to power the field cage. These two HV inputs allow for independent control of the GEM gain and drift velocity. Two CAEN R1470ETD $\SI{8}{kV}$ rated 8 channel power supplies are used to deliver HV power to the TPCs in the BWD and FWD tunnels.  Two custom HV cables are constructed for each TPC: one cable, terminated with identical SHV plugs on each end, that provides $\SI{2.1}{kV}$ to the GEM voltage divider circuit, and another cable, terminated with an SHV plug on one end and a \SI{10}{kV} rated UHV plug on the other end, to provide up to $\SI{8.0}{kV}$ to the field cage. Remote control of each HV channel is provided using CAEN's proprietary GECO2020 software.

\section{Simulation}
\label{sec:sim}

A custom, multi-step Monte-Carlo (MC) simulation pipeline is utilized to produce files containing recoils in each TPC, which can be compared directly with experimentally measured recoils. We describe the production steps here:

\begin{enumerate}
\item \textbf{Generation of beam background events}: The tools used to simulate the production of background particles differ depending on the background source. For Bremsstrahlung, Coulomb, and Touschek backgrounds--those generated from circulating single beams--particle scattering and loss positions are simulated using the Strategic Accelerator Design (SAD) framework \cite{sad}. For luminosity backgrounds, \texttt{FORTRAN} based event generators \cite{Torben} are used to simulate background particles originating from collision-based physics processes. The kinematic information of these initial background particles is saved and passed into Geant4 \cite{geant,geant2,geant3}.

\item \textbf{Simulation of background showers and neutron propagation}: We use a custom ``far beamline" geometry implementation in Geant4 that is integrated into the Belle II Analysis Software Framework (basf2) \cite{basf2, basf22}, which simulates the entire geometry of the SuperKEKB-Belle II system within $|z|<\SI{29}{m}$ from the IP. \autoref{fig:phase3} shows a cross sectional material scan of this region which includes material descriptions of the beam pipes, magnets, collimators, Belle II detectors, and background detectors including BEAST TPCs. Once initial background particles are loaded, Geant4 simulates the physics of the interaction between these particles and the materials present within the geometry. Neutron interactions are simulated using the Geant4 Neutron Data Library (\texttt{G4NDL4.6}) \cite{JEFF}. The positions and momenta of all particles depositing energy within the sensitive volume of each TPC are saved. The initial production positions and momenta of all neutrons depositing energy within a TPC are also saved.

\end{enumerate}

Many improvements have been made recently to the implementation of tracking and recording SuperKEKB losses in SAD \cite{natochii2}, as well as the implementation of accelerator and detector components for SuperKEKB and Belle II in Geant4 \cite{liptak, natochii}.

\subsection{Simulation of TPC detectors}
\label{subsec:sim_description}
\begin{figure}[htbp]
\begin{center}
\includegraphics[width=0.45\textwidth]{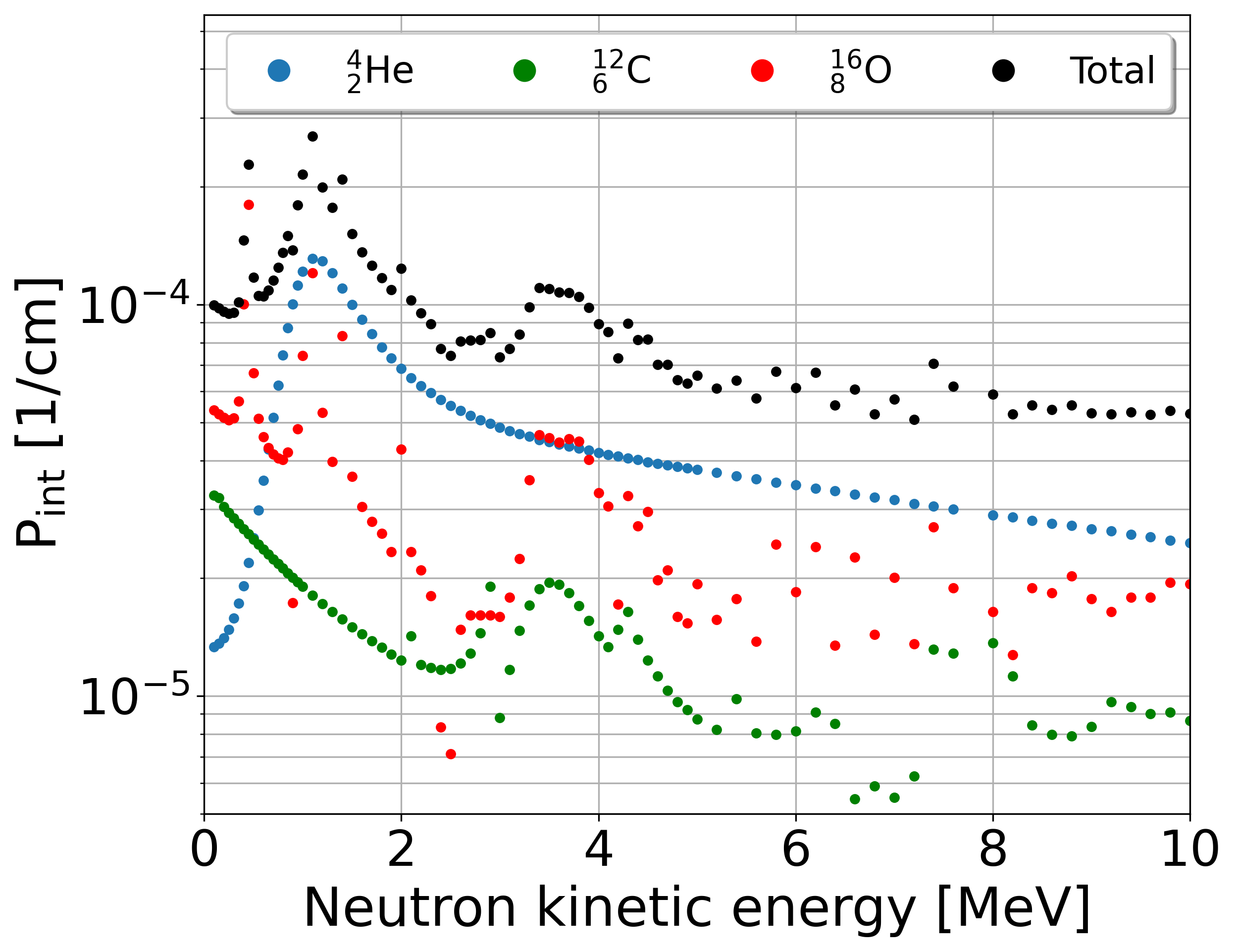}
\caption{(color online) Neutron interaction probability per centimeter ($\text{P}_\text{int}$) of gas traversed versus neutron kinetic energy in a 70:30 mixture of He:CO$_2$ at $\SI{1}{atm}$ pressure. Elastic scattering cross sections to produce this figure were obtained from the \texttt{ENDF/B-VII} database \cite{ENDF} and may differ slightly from those listed in the \texttt{G4NDL4.6} library.}
\label{fig:probs}
\end{center}
\end{figure}

\autoref{fig:probs} shows the probability of an incident neutron interacting with a gas nucleus per centimeter of travel within the fiducial volume of the vessel. Given the low probability of interaction of a neutron passing through the fiducial volume of a TPC, we scale up the elastic scattering cross sections between neutrons and $^4\text{He}$, $^{12}\text{C}$, and $^{16}\text{O}$ nuclei in the \texttt{G4NDL4.6} library, each by factors of 100, to reduce the computational resources necessary to generate adequate simulated nuclear recoil samples for analysis. This leads to a roughly 100-fold increase in the nuclear recoil detection efficiency in each TPC, so when comparing measured and simulated nuclear recoil rates in the TPCs in \def\sectionautorefname{Section}\autoref{sec:results}, we divide the rates predicted by simulation by 100 to compensate for these cross section adjustments.

The Geant4 simulation step provides energy deposits, positions, and momenta of simulated particles within the sensitive volume of a TPC. For each energy deposit, an ionization distribution is created with the number of electrons in the distribution determined from a random Gaussian distribution centered at $\mu = \frac{E_\text{dep}}{W}$, with spread $\sigma = \sqrt{F\mu}$. Here $W$ is the work function (average energy per electron-ion pair) of the gas mixture, $F$ is the Fano factor, and $E_\text{dep}$ is the ionization energy of the energy deposit. $W$ is determined to be $\SI{34.45}{eV}$ using \texttt{Garfield++} \cite{garfield} and $F$ is set to 0.19, as provided by \texttt{Heed}. Each ionization distribution is then read into a TPC simulation framework developed by the authors. 

This custom TPC simulation models the ionization drifting through the simulated field cage volume toward a double GEM layer where the charge is amplified and then ``pixelized" into a $2.00\times \SI{1.68}{cm}^2$ plane containing an $80\times 336$ array of $250\times \SI{50}{\um}^2$ pixels, mimicking the FE-I4B sensitive area. All position resolution effects in the TPC are modeled using Gaussian effective resolutions, as described in \cite{vahsen3d}. Longitudinal and transverse diffusion constants for the drift volume, and for high-field regions internal to the readout plane were estimated using Magboltz \cite{Magboltz}. Each primary electron is diffused individually in three dimensions, based on these diffusion constants and the drift distance to the readout plane. Avalanche gain is simulated at the single-electron level, using an exponential gain distribution. Secondary electrons after gain are again smeared individually, by a combined readout resolution. In the transverse direction, this readout resolution includes contributions from quantization into two GEM holes, diffusion in the transfer gap between the GEMs, and diffusion in the collection gap between the bottom GEM and the pixel chip. Longitudinal diffusion in the same two regions is also included. Finally, the charge is quantized into FE-I4B pixels and readout time bins, and converted into pixel chip-specific charge units called time over threshold (TOT; see \def\sectionautorefname{Section}\autoref{sec:calibration}), based on TPC-specific chip calibration configurations which mimic the experimentally determined charge calibration to convert TOT into charge. These simulated quantized charge distributions encode the same 3D information present in a measured event.

\subsection{Description of simulated samples generated for analysis} 
\label{subsec:MC_description}
Separate samples for Coulomb, Bremsstrahlung, and Touschek scattering were generated in each ring assuming beam optics and machine parameters representative of conditions recorded during Study A. The number of initial simulated particles passed into Geant4 from these single-beam sources is determined from the loss distributions computed by SAD. The effects of beam pipe pressure on beam-gas losses in SAD are weighted based on measurements from over 300 cold cathode gauges (CCGs) spread around each ring. This implementation \cite{natochii} provides a more realistic modeling of the beam pipe gas pressure than assuming uniform pressure distributions. Base and dynamic pressure contributions to the SAD simulated Coulomb and Bremsstrahlung losses are separated into individual components, with the base pressure sample assuming $P(I=\SI{0}{A})$ and the dynamic pressure sample assuming $P(I=\SI{1.2}{A})-P(I=\SI{0}{A})$. Machine parameters used in determining losses are summarized in \autoref{tab:sim_breakdown}. Collimator tip scattering and  a more accurate beam pipe shape were also implemented in SAD \cite{natochii2}, giving better confidence in the modeling of losses from circulating beams around the ring.
\begin{table}[htbp]
\begin{center}
\begin{tabular}{ccccc}
\toprule
 & I[A] & $\sigma_y$[\SI{}{\um}] & $n_b$ & Luminosity [cm$^{-2}$s$^{-1}$]   \\ \hline
LER &  1.2 & 37  & 1576 &  \multirow{2}{*}{$2.5\times 10^{35}$} \\
HER  &  1.0 & 36 & 1576 \\
\bottomrule
\end{tabular}
\caption{Accelerator and luminosity conditions used to generate MC events.}
\label{tab:sim_breakdown}
\begin{tabular}{cc}
\toprule
\begin{tabular}[c]{@{}c@{}} Background \\ Type \end{tabular} & \begin{tabular}[c]{@{}c@{}}Simulated \\ Beam Time [s] \\ (LER,HER) \end{tabular}\\ \hline
Coulomb & (4,40) \\
Bremsstrahlung & (40,400) \\
Touschek & (0.4,1.6) \\ 
Radiative Bhabha & 0.0097 \\
Two-Photon & 0.01 \\
\bottomrule
\end{tabular}
\caption{Total simulated beam time for each background process. Values within the parenthetical numerical pairs denote the beam time of single beam simulation samples in each ring.}
\label{tab:MC_sample}
\end{center}
\end{table}
\begin{table*}[htbp]
\begin{center}
\begin{tabular}{cccccccc}
\toprule
TPC location & $Q_\text{TOT=0}$ [e$^-$] & $Q_\text{TOT=13}$ [e$^-$] & $G$ [e$^-$] & $E_\text{drift}\left[\frac{\text{V}}{\text{cm}}\right]$ & $v_d\left[\frac{\SI{}{\um}}{\SI{25}{ns}}\right]$ & \begin{tabular}[c]{@{}c@{}}Correction\\ Template \end{tabular} & Threshold [$\rm keV_{ee}$]\\ \hline
$z = \SI{-14}{m}$   & $2120 \pm 46$ & 42827 & $783 \pm 45$ & 313 & 152 & $f_3$ & 8.0\\
$z = \SI{-8.0}{m}$  & $2773 \pm 40$ & 47989 & $794 \pm 37$ & 452 & 216 & $f_3$ & 6.0\\
$z = \SI{-5.6}{m}$  & $2110 \pm 32$ & 40821 & $1015 \pm 42$ & 452 & 216 & $f_1$ & 9.5\\
$z = \SI{+6.6}{m}$  & $2071 \pm 50$ & 46794 & $1476 \pm 27$ & 452 & 216 & $f_1$  & 10\\
$z = \SI{+14}{m}$  & $2084 \pm 35$ & 47304 & $883 \pm 34$ & 452 & 216 & $f_2$  & 8.0\\
$z = \SI{+16}{m}$ & $2083 \pm 45$ & 43625 & $863 \pm 84$ & 452 & 216 & $f_3$   & 10\\
\bottomrule
\end{tabular}
\caption{TPC calibration results and settings. Columns from left to right describe: (i) TPC location in Belle II coordinate system, (ii) measured threshold charge, (iii) measured saturation charge, (iv) calibrated effective double GEM gains, (v) approximate drift field, (vi) approximate drift speed, (vii) best template model used for correcting energy due to gain drops during background studies, and (viii) the determined X-ray veto threshold using the procedure outlined in \def\subsectionautorefname{Section} \autoref{subsec:pid}. Threshold uncertainties (ii) are the standard deviation of the pixel threshold over every pixel in the chip. Gain uncertainties (iv) are given by the standard error of the ionization energy distributions of horizontal alphas in a given TPC. Drift speeds (vi) are calculated using Magboltz \cite{Magboltz}. We note that the TPC at $z = \SI{-14}{m}$ experienced periodic high voltage trips, so its drift field was lowered to ensure stable operation.}
\label{tab:calibration}
\end{center}
\end{table*}
For luminosity background samples, we found that the overwhelming majority of simulated radiative Bhabha background events come from the \texttt{BBBREM} event generator \cite{Kleiss} (events leading to neutron recoils in the TPCs from the \texttt{BHWIDE} generator \cite{Jadach} were negligible and are thus not included in this analysis). Two-photon background samples are generated using the \texttt{AAFH} event generator \cite{Berends} with the final state set to $e^+e^-e^+e^-$. Both the radiative Bhabha and two-photon neutron background samples are generated assuming a luminosity of $\SI{2.5e35}{cm}^{-2}$s$^{-1}$. Given the linear dependence of collision rates on luminosity, we can scale these rates to any luminosity as needed. \autoref{tab:MC_sample} summarizes the amount of beam time used for each simulated background component.

\section{Detector calibration and event selection}
\label{sec:calibration}

In this section we give a brief overview of the event-level data processing for each TPC and describe the steps taken to calibrate charge, gain, and ultimately determine particle identification (PID) criteria in each TPC. More detailed information about charge readout and general calibration procedures for the FE-I4B readout chips in these TPCs can be found in Ref. \cite{Jaegle}, and more general information about ATLAS FE-I4B performance and calibration can be found in Ref. \cite{Ahlburg}.

\subsection{TPC data processing and pixel-level calibration}

In both TPC DAQ systems, we utilize firmware packaged with the \texttt{pyBAR} readout software  \cite{pybar} that enables asynchronous triggering of each pixel chip. We collect data using a custom self-triggering setting that provides a readout time window of $\SI{2.5}{\us}$ per event \cite{Jaegle}.

Each pixel cell in a given FE-I4B contains a two-stage charge sensitive integrating amplifier, followed by a comparitor with a set threshold voltage that samples on a \SI{40}{MHz} clock. The comparitor outputs a logical high whenever the signal from the integrating amplifier is above the threshold voltage. The time over threshold (TOT) is the amount of time the comparitor signal outputs logical high and thus depends on the amount of charge deposited on the pixel cell.

The first charge above threshold in an event triggers the readout. When an event is triggered, pixel hit data such as the column and row of the ($80 \times 336$) pixel matrix, the TOT, and the readout time (recorded in multiples of \SI{25}{ns} due to the \SI{40}{MHz} comparitor clock) are recorded. Given that ionization drifts, on average, at a constant speed $v_d$ within the sensitive volume of a TPC, the readout time is used to construct a relative $z$ coordinate 

\begin{align}
z_{rel,\text{TPC}} = v_dn\times\SI{25}{ns},
\end{align} 
where $n\times\SI{25}{ns}$ is the readout time within the $\SI{2.5}{\us}$ event readout window. For the purposes of this work, we only need event timestamps to correlate our TPC measurements with SuperKEKB machine measurements that update every second, so we use a software-level trigger-counter shared by all TPCs, which updates on a \SI{20}{Hz} clock.

The charge read out in an FE-I4 pixel can be calibrated by first tuning the comparitor threshold voltage and the TOT. A charge injection circuit is located in each pixel which injects charge in discrete voltage steps through two injection capacitors; these injection capacitors make the injected charge proportional to a variable injection voltage, thereby creating charge pulses of different, known, magnitude. The \texttt{pyBAR} readout software is packaged with a global tuning script that iteratively tunes both the threshold and TOT response of the integrating amplifier for all 26,880 pixels in the chip. The thresholds are tuned to a target value of \SI{2100}{e} in all but one of the TPCs and the TOT response is tuned to correspond to a saturation limit--the maximum TOT recorded in an event--above \SI{40000}{e}. Due to operational oversight, the TPC at $z=\SI{-8.0}{m}$ was tuned to a target threshold of \SI{2750}{e} instead of \SI{2100}{e}. The results of the threshold tunings and TOT scales vary between FE-I4 modules as can be seen in \autoref{tab:calibration}. The measured TOT in given pixel hit is a 4 bit integer code ranging from 0 to 13, with TOT = 0 corresponding to a pixel near threshold and TOT = 13 corresponding to a saturated pixel.

After tuning the threshold and the TOT response in each TPC, we calibrate the remainder of the charge scale by sending two hundred injections at each of several distinct charge steps in each FE-I4B pixel and measure the TOT response in each pixel. The mean of the TOT over all pixels is recorded for each injection and the mean and standard error of each of these 200 pixel-averaged TOT values is plotted at each charge step. A bi-cubic spline interpolation of the injection charge versus mean TOT is used to determine the charge corresponding to TOT codes ranging between 0 and 13 in steps of 1.
\begin{figure*}[htbp]
\centering
\includegraphics[width=\textwidth]{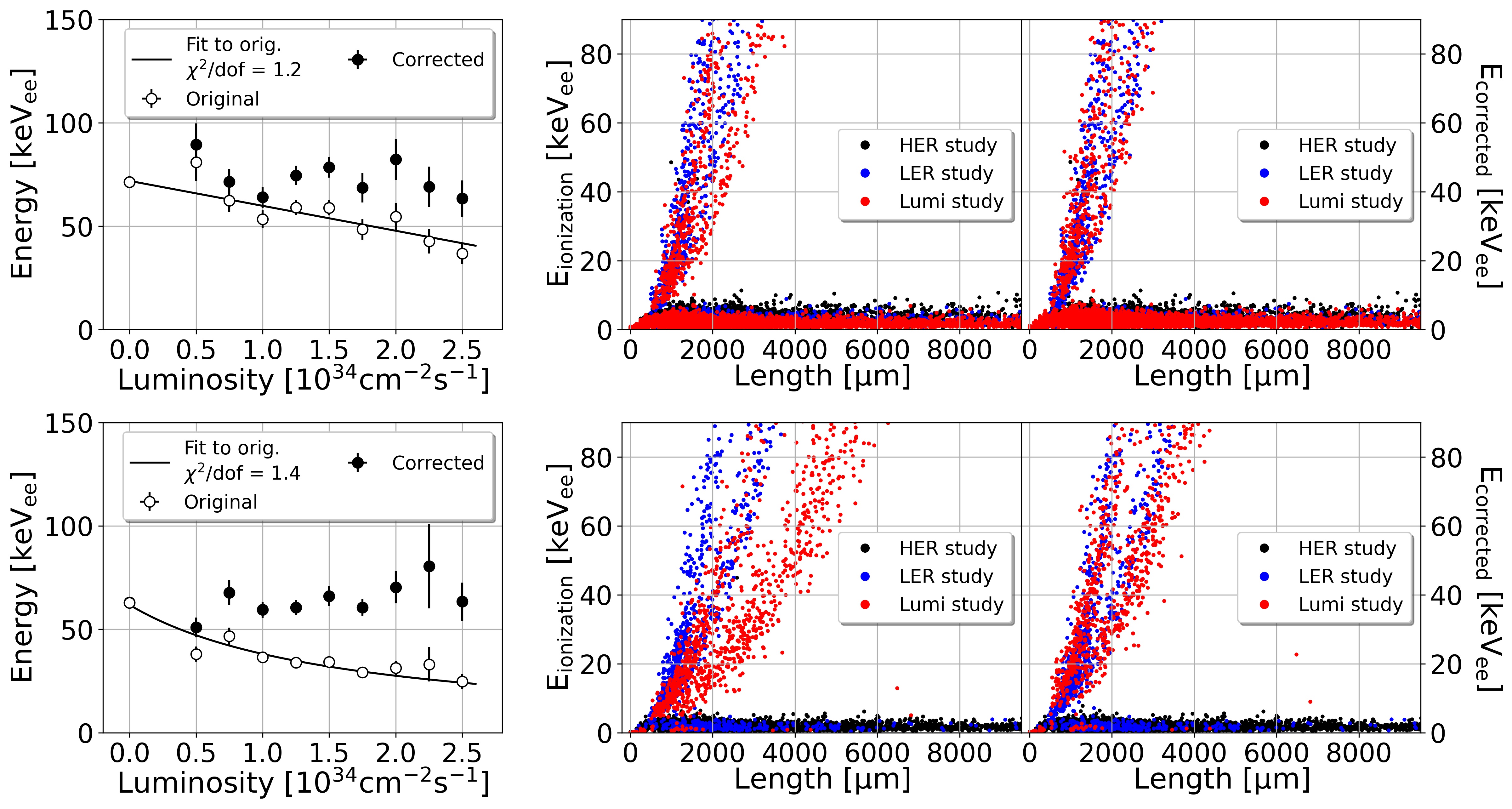}
\caption{(color online) Left: average energy of identified nuclear recoils--both before and after the corrections described in Eqs. (\ref{eq:cor1}) and (\ref{eq:cor2})--versus luminosity for recoils satisfying $\SI{1.3}{mm} < \ell < \SI{2.6}{mm}$ (top) and $\SI{1.7}{mm} < \ell < \SI{3.0}{mm}$ (bottom), where $\ell$ is the length along the principal axis of the recoil track. Middle and right: reconstructed ionization energy versus length distributions before and after gain corrections, respectively. The top and bottom rows of plots are associated with the TPCs at $z = +\SI{14}{m}$ and $z = +\SI{16}{m}$, respectively.}
\label{fig:gain_corr}
\end{figure*}
\subsection{TPC gain calibration}
\label{subsec:gain}
We calibrate the effective gain of each TPC using alpha particles emitted from the $^{210}\text{Po}$ disk source installed in each detector (see \autoref{fig:tpc}). The ionization energy from an event in a TPC is related to the observed avalanche charge, $Q$, and effective gain of the TPC, $G$, via

\begin{align}
E_\text{ionization}=\frac{QW}{G},
\end{align}
where $W = \SI{34.45}{eV}$ is the average energy per electron-ion pair of 70:30 He:CO$_2$. The location of the disk source is identical in each TPC and is chosen so that alpha tracks emitted from the $^{210}\text{Po}$ source span the entire $\SI{2}{cm}$ extent of $x_\text{TPC}$. This means the ionization energy deposited on the chip by an alpha with $(\phi_\text{TPC},\theta_\text{TPC})\sim (0^{\circ},90^{\circ})$ will be constant within statistical fluctuations and can thus reliably be used as a reference value to determine $G$.

To ensure a pure sample of alphas, we calibrate using only alpha tracks that were recorded when beams were not circulating the main ring. Simulation shows $\langle E_\text{ionization}\rangle = \SI{1430}{keV}$ for an alpha track with $(\phi_\text{TPC},\theta_\text{TPC}) = (0^{\circ},90^{\circ})$ in 70:30 He:CO$_2$, so we select alphas satisfying $|\phi_\text{TPC}|<5^\circ$ and $85^\circ<\theta_\text{TPC}<95^\circ$ and expect the ionization energy distributions of these to be approximately Gaussian distributed. We thus plot these distributions and set our calibrated gain, $G$, to be the value that centers the ionization distribution of these selected alphas at $\langle E_\text{ionization}\rangle = \SI{1430}{keV}$ in each TPC. The calibrated GEM gains are shown in \autoref{tab:calibration}. We note that effective gain variations between TPCs result both from inherent GEM gain differences and variation in gas purity, the latter being expected due to each set of three TPCs receiving gas in series.

During both Study A and Study B, we observe drops in $\text{d}E/\text{d}x$ at a fixed track length, $\ell$, with increasing luminosity in several TPCs. This effect appears to result from drops in effective gain in these detectors during beam-collisions. A change in background particle composition alone would only change the distribution of events within an $E$ versus $\ell$ band, or the relative normalization of the bands, but not the shape of individual bands comprised of a single particle species. Both the reduced d$E$/d$x$ for the observed He-recoils and the disappearance for the X-ray bands (which move below threshold at low gain) support the hypothesis that the effective gain is reduced at the highest luminosities.
\begin{figure*}[htbp]
\includegraphics[width=\textwidth, trim ={1cm 0cm 1cm 1cm}]{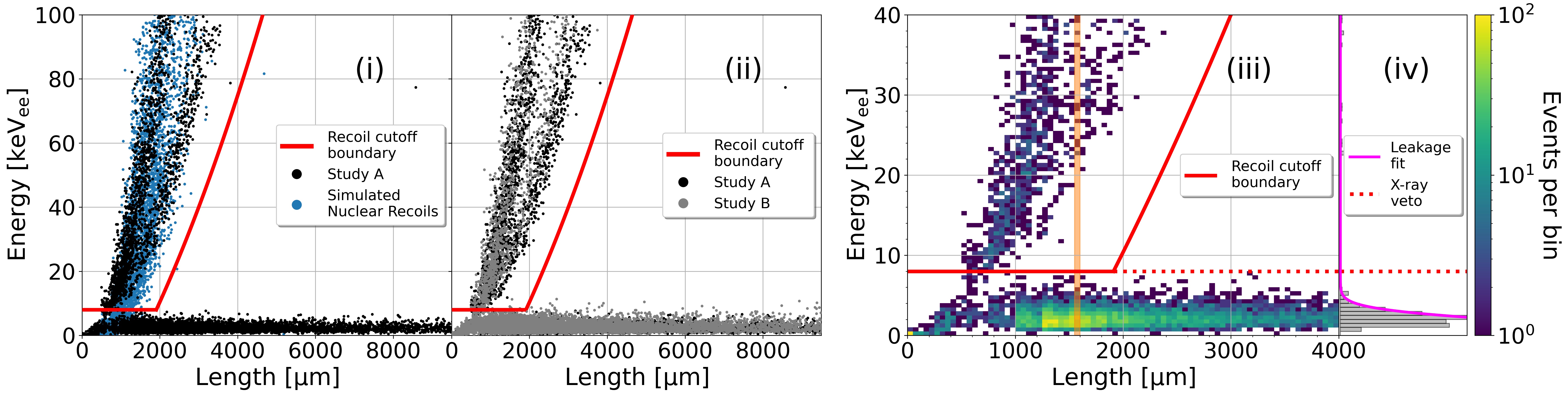}
\caption{(color online) (i): Corrected energy versus 3D track length for simulated nuclear recoil events and all measured events passing our fiducialization preselection during Study A in the TPC at $z_\text{BELLE}=+\SI{14}{m}$. Events above the red cutoff boundary are identified as nuclear recoils, and those below are rejected background. (ii): same as (i) but comparing measurements between Study A (black points) and Study B (gray points). (iii): zoomed in binned view of measured recoils in (i). (iv): binned energy distribution of all events within the orange shaded region shown in (iii). The energy distribution of events below the X-ray veto is fit with a half-Gaussian profile $\mathcal{F}_\text{bin}(x)$ (described in text and shown as the magenta curve here) to estimate the leakage of X-ray events above this threshold.}
\label{fig:evl}
\end{figure*}

Given that beam-induced background rates increase with luminosity, we speculate that the observed drop in effective gain is a consequence of reduced gas purity induced by stimulated X-ray desorption at higher luminosities. Since this is speculation and we do not know the exact cause of these drops in gain, we correct for them empirically to keep the $\text{d}E/\text{d}x$ distributions of nuclear recoils reasonably consistent in a given TPC during the entirety of the background studies. We thus introduce three simple template models to fit to the distribution of mean nuclear recoil energies for events with observed ionization energies above \SI{12}{keV} with 3D lengths (\def\subsectionautorefname{Section}\autoref{subsec:pid}) between \SI{1.3}{mm} and \SI{2.6}{mm} (\SI{1.7}{mm} and \SI{3.0}{mm} for the TPC at $z=+\SI{16}{m}$ due to its sharper drop in effective gain) binned by luminosity and choose the model among the three that gives a reduced $\chi^2$ nearest to 1. The template models are

\begin{align}
\label{eq:cor1}
f_1(L) &= a \nonumber \\
f_2(L) &= aL+b \nonumber \\
f_3(L) &= \frac{1}{aL+b}
\end{align}
where $L$ is the measured luminosity and $a$ and $b$ are fit parameters. We then use $f_j(0)/f_j(L);\text{ } j=1,2,\text{ or } 3$, as the scale factor for which to modify the energy. Thus, for all events, we define our corrected energy due to gain corrections as

\begin{align}
\label{eq:cor2}
E_{\text{corrected}}(L) = \frac{f_j(0)}{f_j(L)}E_{\text{ionization}},
\end{align}
where $j$ corresponds to the template that minimizes $|\chi^2/\text{dof} - 1|$. \autoref{tab:calibration} lists which fit template was used to correct for drops in gain in each TPC and \autoref{fig:gain_corr} shows the effect of these corrections on the TPCs at $z=+\SI{14}{m}$ and $z=+\SI{16}{m}$. Moving forward, unless stated otherwise, when we refer to the energy of a measured recoil event, we are referring to $E_\text{corrected}$.

\subsection{Event classification}
\label{subsec:pid}

The ionization energy of a track and its length in three dimensions provide sufficient information for selecting high purity nuclear recoil samples in a TPC. Once charge and gain have been calibrated, the ionization energy of a track is measured by summing over the energy deposited in each pixel hit in the event

\begin{align}
E_\text{ionization} = \sum_\text{hits}E_\text{hit}.
\end{align}

Using a Singular Value Decomposition (SVD), we identify the principal axis of each track and take the difference between the highest and lowest values of position along this axis to be the 3D track length. Before determining nuclear recoil selections, we apply a ``fiducialization preselection" where we reject all events that register pixel hits on any of the four edges of the FE-I4B chip. This removes calibration alpha tracks since they span the entire $x_\text{TPC}$ extent of the chip and also helps ensure that the recorded ionization energy isn't biased by events with charge outside of the fiducial volume of the TPC.

\autoref{fig:evl} shows distributions of track energy versus length after the fiducialization preselection. In each plot, there are three distinct $\text{d}E/\text{d}x$ bands. The relatively flat lowest energy band ($\text{d}E/\text{d}x\sim 0$) corresponds to electron recoils from X-ray conversions, which is the predominant source of background in all TPCs. The remaining two curved $\text{d}E/\text{d}x$ bands correspond to, in order of increasing $\text{d}E/\text{d}x$: $^4$He recoils and $^{12}$C/$^{16}$O recoils. The $\text{d}E/\text{d}x$ distributions of recoiling $^{12}$C and $^{16}$O nuclei are similar enough that we do not distinguish between them in measurement. The large region of parameter space between the X-ray and recoil bands contains a relatively small number of events compared to the three primary $\text{d}E/\text{d}x$ bands. Simulation suggests that this region contains a mixture of nuclear recoils and X-rays, but given that there are very few events in this region, we apply quadratic $\text{d}E/\text{d}x$ pre-selections that reject most of this region in order to prioritize nuclear recoil purity. After setting these $\text{d}E/\text{d}x$ pre-selections, we use a data driven approach to determine an ``X-ray veto threshold" in each TPC for our analysis. The general procedure follows:

\begin{enumerate}
\item Start with an X-ray veto threshold of \SI{5}{keV_{ee}}.
\item Split the region between $\ell = \SI{850}{\um}$ and $\ell = \SI{4000}{\um}$ into bins of width $\SI{50}{\um}$ and in each of these length bins:
\begin{enumerate}
\item Plot a histogram of the energy distribution of all events. The gray bars in \autoref{fig:evl} (iv) show this energy distribution for all events within the length bin shaded in orange in \autoref{fig:evl} (iii). The distribution of events below the recoil cutoff boundary is asymmetric and has a long, approximately Gaussian, high-energy tail.
\item Estimate the number of x-ray events as a function of energy for a given length bin using a Gaussian profile of the form

\begin{align}
\mathcal{F}_\text{bin}(E_\text{corrected}) = \hat{A}_\text{bin}\exp(-\hat{B}_\text{bin}E_\text{corrected}^2),
\end{align}

fit to the higher-energy half of the observed energy distribution below the recoil cutoff boundary. Fit parameters $\hat{A}_\text{bin}$ and $\hat{B}_\text{bin}$ are determined by performing a $\chi^2$ minimization of $\mathcal{F}_\text{bin}$ evaluated between the energy bin with the most X-ray events and bin corresponding to the recoil cutoff boundary in the given length bin.
\item Estimate the leakage above the recoil cutoff boundary using

\begin{align}
\label{eq:leakage}
\text{Leakage}_\text{bin} = \int_{\max_E\left(E_\text{veto}, E_{\text{d}E/\text{d}x,\text{bin}}\right)}^\infty \mathcal{F}_\text{bin}(x)dx,
\end{align}

where $E_\text{veto}$ and $E_{\text{d}E/\text{d}x,\text{bin}}$ are the energies of the proposed X-ray veto threshold and the d$E$/d$x$ pre-selection, respectively.
\item Divide the estimated leakage above the recoil cutoff boundary by the total number of events above $\max_E\left(E_\text{veto}, E_{\text{d}E/\text{d}x,\text{bin}}\right)$ to get the estimated \textit{leakage fraction} above $\max_E\left(E_\text{veto}, E_{\text{d}E/\text{d}x,\text{bin}}\right)$. Subtracting this leakage fraction from 1 gives the estimated recoil purity above the recoil cutoff boundary.
\end{enumerate}
\item Adjust the energy of this flat X-ray veto threshold as needed until the estimated recoil purity in each length bin is greater than $99\%$.
\end{enumerate} 

We apply the above procedure to each TPC twice--once for Study A data and once for Study B data--and note that the determined thresholds for a given TPC are similar between these two studies, so we assign the larger of the two thresholds to each TPC with the final choice listed in column (viii) of \autoref{tab:calibration}. These thresholds are defined with respect to $E_\text{corrected}$\footnote{We increase the X-ray veto threshold in the TPC located at $z=+\SI{16}{m}$ by an additional $60\%$ to ensure that we are above the regime where nuclear recoil events drop below the FE-I4B detection threshold due to the drop in gain during collisions. The X-ray veto thresholds reported in \autoref{tab:calibration} for this TPC are after the additional $60\%$ increase.}. Despite small differences in $\text{d}E/\text{d}x$ predictions between data and simulation, applying these same selections to simulated samples suggests recoil purities of greater than $99\%$ in all TPCs with nuclear recoil signal efficiencies greater than $75\%$ in all TPCs. The predicted nuclear recoil signal purities and efficiencies demonstrate that these TPCs are capable of measuring high purity samples of fast neutrons down to $\mathcal{O}(\SI{10}{keV})$ at effective double GEM gains of $\mathcal{O}(1000)$.

\subsection{Merging accelerator and TPC data}
Before detailing our analysis methodology, we provide a brief overview of the data processing steps taken to merge accelerator parameters with calibrated TPC measurements. Summaries of key SuperKEKB parameters are stored as process variables using the Experimental Physics and Industrial Control System (EPICS PVs) \cite{EPICS} that update every second. These PVs are archived internally for Belle II and SuperKEKB collaborators using custom software built around the EPICS Archiver Appliance \cite{park}. All accelerator parameters used in this analysis are extracted from this PV archiver and are merged with calibrated TPC data by matching integer timestamps of TPC data with integer timestamps of all accelerator PVs. For cases where there are multiple TPC events within a one second window, the accelerator data is duplicated for each TPC event. In this way, when we model nuclear recoils as a function of accelerator parameters, these models are \textit{rate weighted}.

\section{Modeling beam-induced backgrounds}
\label{sec:Analysis}

\begin{figure*}[htbp]
\begin{center}
\begin{tikzpicture}[ultra thick, overlay]
\draw [pen colour={red},
    decorate,
    decoration = {calligraphic brace,
        raise=1pt,
        amplitude=5pt},xshift=-3cm,yshift=0.6cm] (-3.7,-1) -- (-1,-1) node [black,midway,yshift = 0.5cm]
{\footnotesize \textcolor{red}{\textbf{LER study}}};

\draw [pen colour={blue},
    decorate, 
    decoration = {calligraphic brace,
        raise=1pt,
        amplitude=5pt},xshift=-3cm,yshift=0.6cm] (-0.5,-1) -- (4.7,-1) node [black,midway,yshift = 0.5cm] {\footnotesize \textcolor{blue}{\textbf{HER study}}};
\draw [pen colour={goold},
    decorate, 
    decoration = {calligraphic brace,
        raise=1pt,
        amplitude=5pt},xshift=-3cm,yshift=0.6cm] (5.4,-1) -- (8.7,-1)
        node [black,midway,yshift = 0.5cm] {\footnotesize \textcolor{goold}{\textbf{Luminosity study}}};
        \end{tikzpicture}

\includegraphics[width=\textwidth]{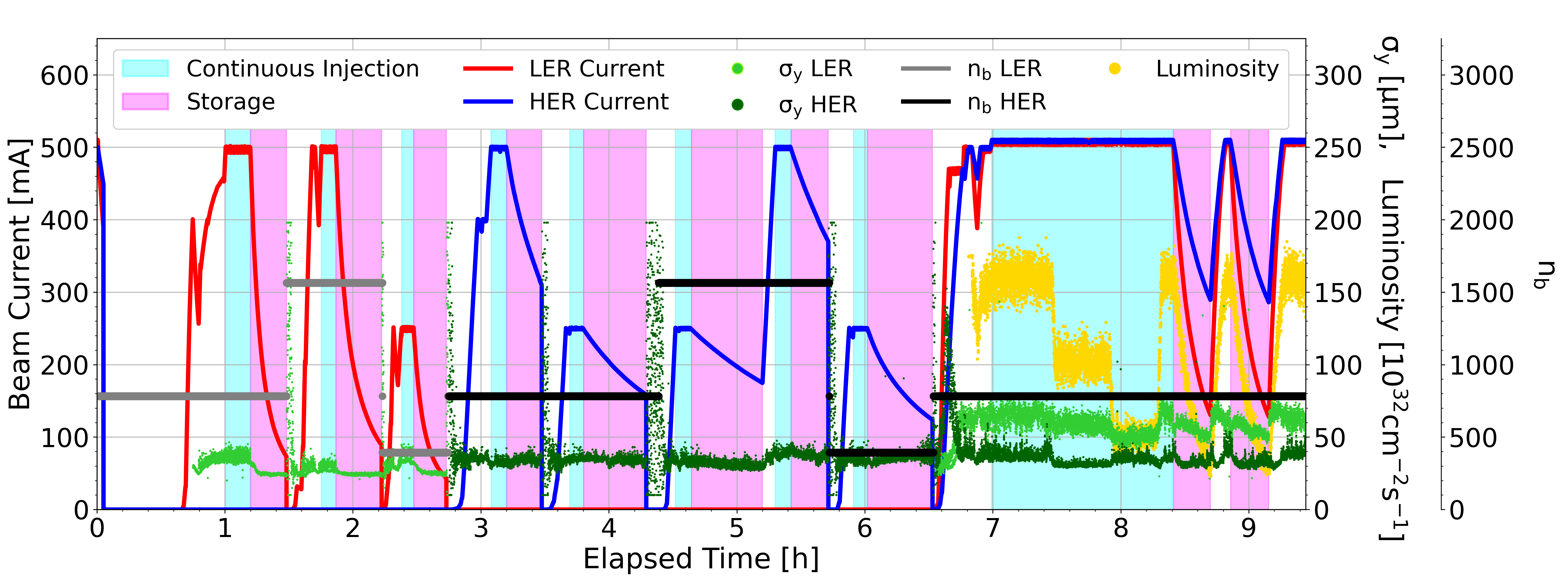}
\caption{(color online) SuperKEKB machine parameters versus time. Beam currents, vertical beam sizes $\sigma_{y,\text{LER,HER}}$, numbers of bunches $n_{b,\text{LER,HER}}$, and luminosity, $L$, are shown over the course of all three study periods during Study A}
\label{fig:summary}
\end{center}
\end{figure*}
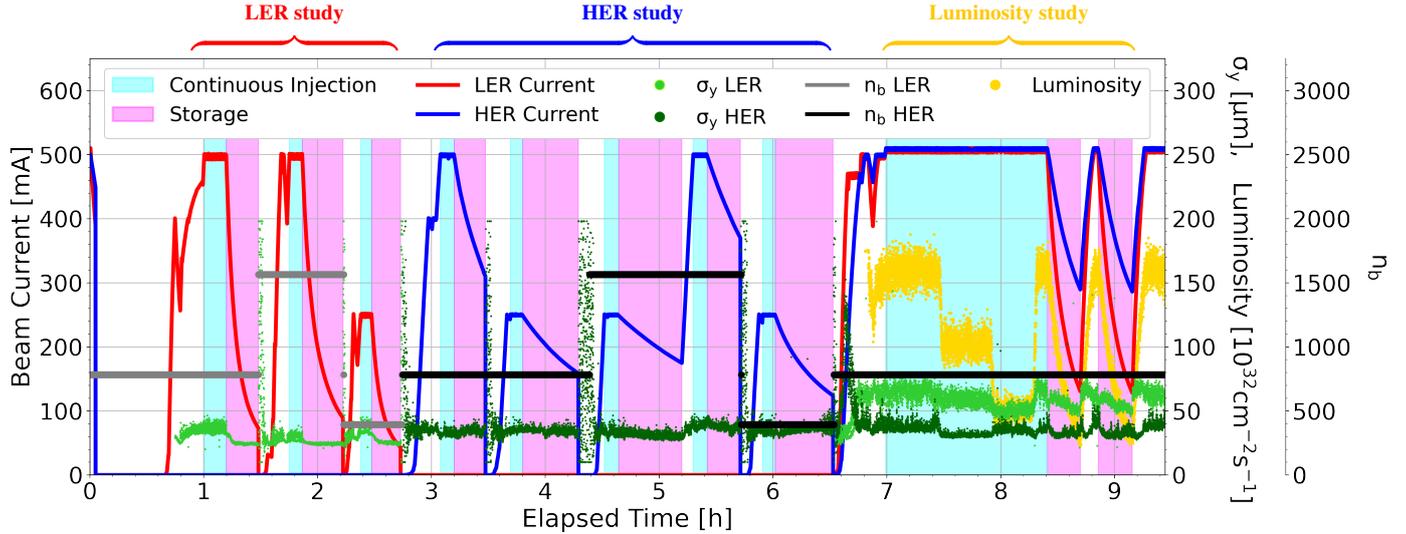

\autoref{fig:summary} shows how several accelerator parameters vary over the course of the three separate study periods conducted during Study A: we see that numbers of bunches ($n_b$) were varied in three steps during each of the LER and HER study periods. Since Touschek backgrounds vary with charge density, explicitly changing the total number of bunches provides a probe for disentangling Touschek background contributions from beam-gas contributions. During the luminosity study period, the number of bunches were kept constant with 783 bunches in each ring, while luminosity was varied to control for the measurement of luminosity dependent backgrounds. During all three background studies, separate continuous injection and beam storage ``decay" fills (cyan and magenta regions in \autoref{fig:summary}, respectively) were collected. When analyzing data, we only consider beam storage fills in our analysis, as there are more systematic differences between the beam optics settings used during during continuous injection fills than those during decay fills.

A chief goal of performing dedicated background studies is to disentangle all contributions to beam-induced background rates. Disentangling background contributions allows for a direct comparison between observed and predicted neutron background compositions, providing validation of our modeling of the mechanisms of neutron background production at SuperKEKB.

\subsection{Beam-gas backgrounds}
\label{subsec:beamgas_bgs}
Beam-gas scattering occurs when $e^+$ and $e^-$ beam particles interact with gas atoms present within the beam pipes. These interactions can either happen through Coulomb scattering between the beam particle and the gas atom, or through Bremsstrahlung. To good approximation, the rate of beam-gas scattering events in a given ring is proportional to $IPZ^2$, where $I$ is the beam current, $P$ is pressure at a given position along the ring, and $Z^2$ is the square of the atomic number of the gas constituent that the beam particles interact with. $P$ is composed of a base component, $P_0$ due to the residual gases remaining in the beam pipe without the presence of beams, and a dynamic component $P_1$, which is proportional to $I$ and $\text{d}P/\text{d}I$ \cite{suetsugu}.

We employ a physically motivated parameterization of nuclear recoil rates, $R$, that encapsulates the sensitivity of $R$ to beam-gas and Touschek background components. Though beam-gas scattering rates scale as $IPZ^2$, there are only 3 residual gas analyzers present around the ring which can provide measurements of the gas composition inside the beam pipe. Since we are unable to directly measure this local gas composition variation throughout the majority of the accelerator rings, we parameterize the beam-gas contributions to nuclear recoil rates, $R_{bg}$ as

\begin{align}
\label{eq:1}
R_{bg} = B\cdot IP,
\end{align}
where we absorb $Z^2$ into the \textit{beam-gas sensitivity} parameter, $B$. We note that $Z$ is taken to be 7 in SAD simulation, so absorbing $Z^2$ into $B$ means that $B$ is not a constant and will in general vary with time and position along the beam pipe.

The pressure in the beam pipe also varies considerably at different locations around the ring, however, these variations are to some degree accounted for in MC through weighting the MC pressure by measured values of CCG pressures. Though there are more than 300 CCGs around the rings, the average spacing between CCGs is $\mathcal{O}(\SI{10}{m})$, so the CCG placement is sparse compared to the expected spatial scale of pressure variations. Despite this, the MC pressure reweighting is still an improvement over the treatment of pressure in previous analyses where we simulated a uniform pressure distribution \cite{lewis, liptak}. 

In experiment, we treat pressure as follows: we first note that the pressure within the beam pipe, $P$, contains both a base component, $P_0$, and a current-dependent dynamic component, $P_1(I)$:

\begin{align}
\label{eq:2}
P &= P_{0} + \frac{\text{d}P}{\text{d}I}\cdot I \nonumber \\
&= P_{0} + P_{1}(I).
\end{align} 
Previous machine simulation has shown that the dynamic pressure component measured by the CCGs, $P_{1,meas}(I)$ is about a factor of 3 lower than the dynamic pressure within the beam pipe, so

\begin{align}
\label{eq:3}
P_{1}(I) &= 3P_{1,meas}(I), \nonumber \\
P_{0} &= P_{0,meas}, \\ \nonumber
&\Rightarrow P = P_{0,meas} + 3P_{1,meas}(I).
\end{align}
The measured CCG pressure can be broken down into base and dynamic components analogously to the beam pipe pressure in Eq. (\ref{eq:2}) so we can rewrite Eq. (\ref{eq:3}) as

\begin{align}
\label{eq:4}
P &= P_{0,meas} + 3(P_{meas}-P_{0,meas}) \nonumber \\
&= 3P_{meas}-2P_{0,meas}.
\end{align}
Substituting Eq. (\ref{eq:4}) into Eq. (\ref{eq:1}) gives

\begin{align*}
R_{bg} = B\cdot I(3P_{meas}-2P_{0,meas}),
\end{align*}
noting that the base pressure is a constant, we define model parameters $B_0\equiv 3B$ and $B_1\equiv 2BP_{0,meas}$ leaving us with our beam gas background parametrization:

\begin{align}
\label{eq:bg}
R_{bg,i} = B_{0i}\cdot I_iP_{meas,i} - B_{1i}\cdot I_i\quad i=\text{LER, HER}.
\end{align}
Cast in the form of Eq. (\ref{eq:bg}), $B_0$ and $B_1$ are positive constants that are determined empirically and together encode the sensitivity of nuclear recoil rates to beam gas backgrounds.
\subsection{Touschek backgrounds}
\label{subsec:single_beam_bgs}

The Touschek effect describes Coulomb scattering between particles within an individual beam bunch, causing the momenta--and by extension, the orbits--of the Touschek scattered particles to deviate from those of the rest of the bunch \cite{piwinski}. For a single beam particle within a bunch, the Touschek scattering rate is proportional to the particle density of the bunch, $I_b/(\sigma_x\sigma_y\sigma_z)$, where $I_b$ is the bunch current, $\sigma_z$ is the longitudinal bunch width, and $\sigma_x$ and $\sigma_y$ are the horizontal and vertical transverse bunch widths. To model the Touschek background rates around the ring, we multiply the Touschek scattering rate of a single particle by number beam particles around the ring, $I_bn_b$, where $n_b$ is the number of bunches in the beam train, leaving us with Touschek background rates, $R_T$, scaling as

\begin{align}
\label{eq:Touschek}
R_T \propto \frac{I_b^2n_b}{\sigma_x\sigma_y\sigma_z},
\end{align}
which suggests Touschek scattering rates, $R_T$, can be parametrized as

\begin{align}
\label{eq:touschek2}
R_T = T\cdot \frac{I_b^2n_b}{\sigma_x\sigma_y\sigma_z},
\end{align}
where $T$ is the experimental \textit{Touschek sensitivity} parameter, which encodes all effects contributing to Touschek background rates that aren't explicitly adjusted in the experiment. In the analyses that follow, we measure $T$ for nuclear recoils observed in the TPCs, so moving forward, $R_T$ is understood to be the number of Touschek scattering-induced nuclear recoils measured by a TPC. Variations in $\sigma_y$ in the LER can be seen between the LER study period and the luminosity study period (\autoref{fig:summary}). This variation is due to a ``beam blowup" effect that occurs during collisions that can have a strong effect on observed rates, especially in the FWD tunnel where LER Touschek backgrounds from upstream collimators are particularly large. Due to this sensitivity to Touschek backgrounds from upstream collimators, we modify the vertical beam size dependence in our \textit{LER} Touschek model from $\sigma_y \rightarrow \sigma_y^{\alpha_\text{LER}}$, where the exponent $\alpha_\text{LER}$ is determined empirically and corrects for discrepancies between Touschek predictions derived from single beam LER studies and those during the luminosity study period. The assumption underlying this correction is that beam size differences can cause larger variations in Touschek backgrounds for detectors that are sensitive to backgrounds from collimators than those that are not.

We also note that $\sigma_x$ was not well measured and isn't expected to vary strongly, so we absorb the effects of $\sigma_x$ into Touschek sensitivity parameter $T$. We can thus write our Touschek parametrization as

\begin{align}
\label{eq:touschek3}
R_{T,i} = T_i\cdot \frac{I_i^2}{\sigma_{yi}^{\alpha_i}\sigma_{zi} n_{bi}}\quad i=\text{LER, HER},
\end{align}
where we have used $I_b=I/n_b$ to express the bunch current in terms of beam current $I$ and number of bunches $n_b$.
Since we restrict our analyses to beam storage fills, beam-gas and Touschek contributions will be the dominant single beam background sources so following the lead of Ref. \cite{lewis}, we can combine Eqs. (\ref{eq:bg}) and (\ref{eq:touschek3}) and write

\begin{align}
\label{eq:heuristic}
R_\text{SB,i} &= R_{bg,i}+R_{T,i} \\
&= B_{0i}\cdot I_iP_{meas,i} - B_{1i}\cdot I_i + T_i\cdot\frac{I_i^2}{\sigma_{yi}^{\alpha_i}\sigma_{zi}n_{bi}} \quad i=\text{LER, HER} \nonumber,
\end{align}
as our \textit{combined single beam background parametrization}. Fitting measured nuclear recoil data with this parametrization provides empirical measurements of $B_{0i}$, $B_{1i}$, and $T_i$, which can then be compared with simulation and used to extrapolate expected single beam-induced TPC event rates to different accelerator conditions.

\subsection{Luminosity backgrounds}
\label{subsec:rbb}
Photons from radiative Bhabha scattering events are the principal collision-based cavern neutron background production source. The rate of RBB events is proportional to the $e^+e^-$ collision rate, and thus, luminosity. RBB photons will travel predominantly in the $z_\text{BELLE}$ direction and will ultimately collide with the beam pipe walls near where the beam pipes start to curve. Neutrons result from these RBB photon collisions via the Giant Dipole Resonance and are expected to propagate in all directions.

We model luminosity backgrounds as the backgrounds that remain during collisions after subtracting out LER and HER single beam backgrounds. Total nuclear recoil rates, then, are given by

\begin{align}
R = R_{\text{SB,LER}} + R_{\text{SB,HER}} + R_L,
\end{align}
where $R$ is the total nuclear recoil rate and $R_L$ is the luminosity component of nuclear recoil rates. Thus, during the luminosity study period, we measure the luminosity background rate as

\begin{align}
\label{eq:lumi}
R_L = R - \left(\sum_{i=\text{LER,HER}}B_{0i}I_iP_{meas,i} - B_{1i}I_i + T_i\frac{I^2_i}{\sigma_{yi}^{\alpha_i}\sigma_{zi}n_{bi}}\right) = m_LL,
\end{align}
where $L$ is the instantaneous luminosity, and $B_{0i}, B_{1i}$ and $T_i$ represent the sensitivity coefficients measured in experiment from the single beam LER and HER background studies, respectively. $R_L$ is expected to be directly proportional to $L$--which we verify in \autoref{sec:results}--with \textit{luminosity background sensitivity} parameter $m_L$.
\label{subsec:lumi}

\section{Analysis and results}
\label{sec:results}
Unless stated otherwise, all TPC results that follow are for nuclear recoils that are selected following the criteria outlined in \def\sectionautorefname{Section}\autoref{sec:calibration}.
\begin{figure}[htbp]
\begin{center}
\includegraphics[width=0.5\textwidth, trim ={0cm 0cm 0cm 0cm}]{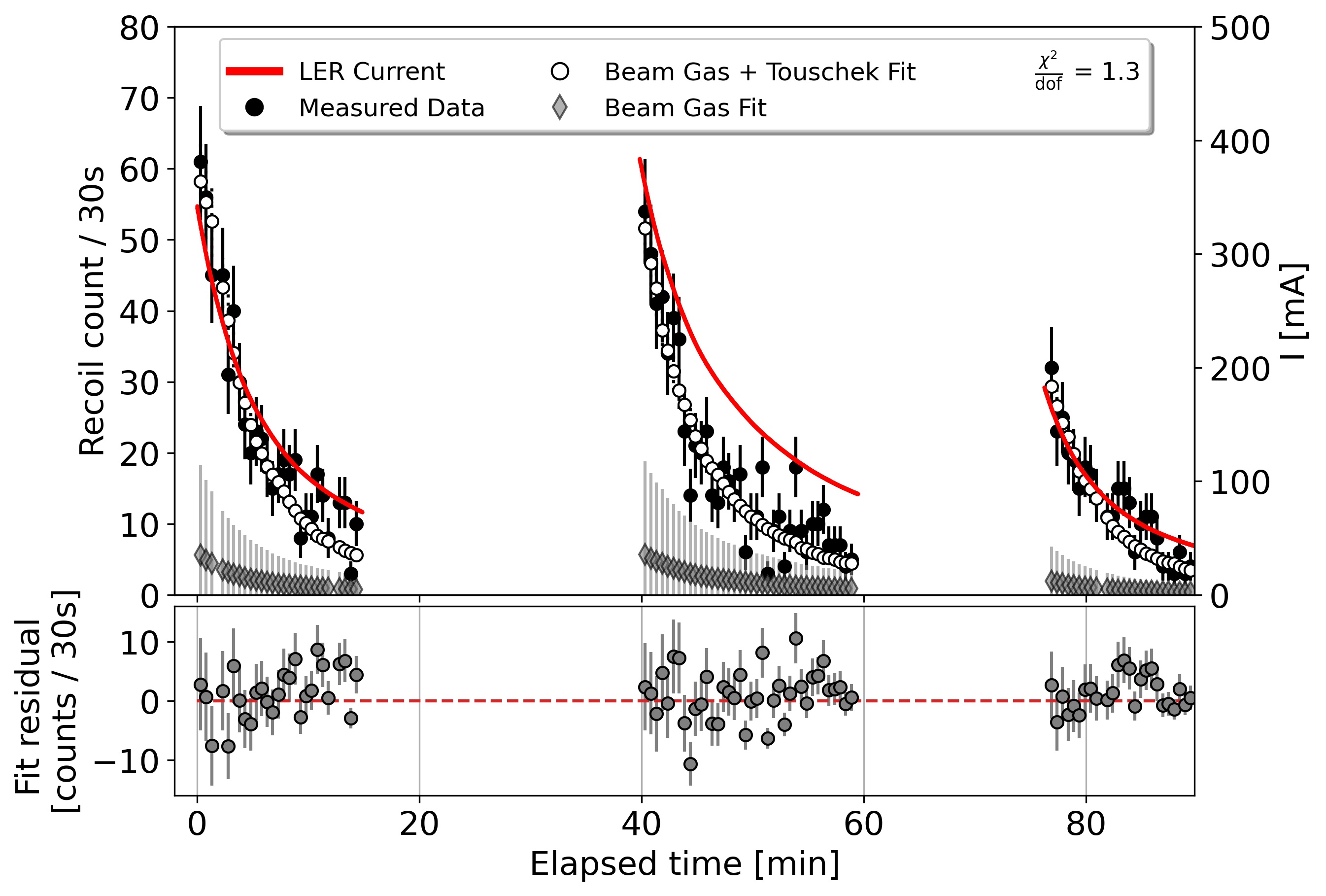}
  \caption{(color online) LER background composition fits versus time in the TPC located at $z_\text{BELLE} = +\SI{14}{m}$ during Study A with $\alpha_\text{LER} = 2.4$. Top: black circles correspond to the measured recoil count, gray diamonds, and open circles represent the predicted recoil counts from LER beam gas and LER beam gas + LER Touschek backgrounds, respectively. Bottom: residual distribution defined as the difference between the measured recoil count and the fit-predicted recoil count using Eq. (\ref{eq:heuristic}).}
  \label{fig:ler_time_fits}
 \end{center}
\end{figure}

\begin{figure*}[htbp]
\begin{center}
\includegraphics[width=\textwidth, trim ={0cm 0cm 0cm 0cm}]{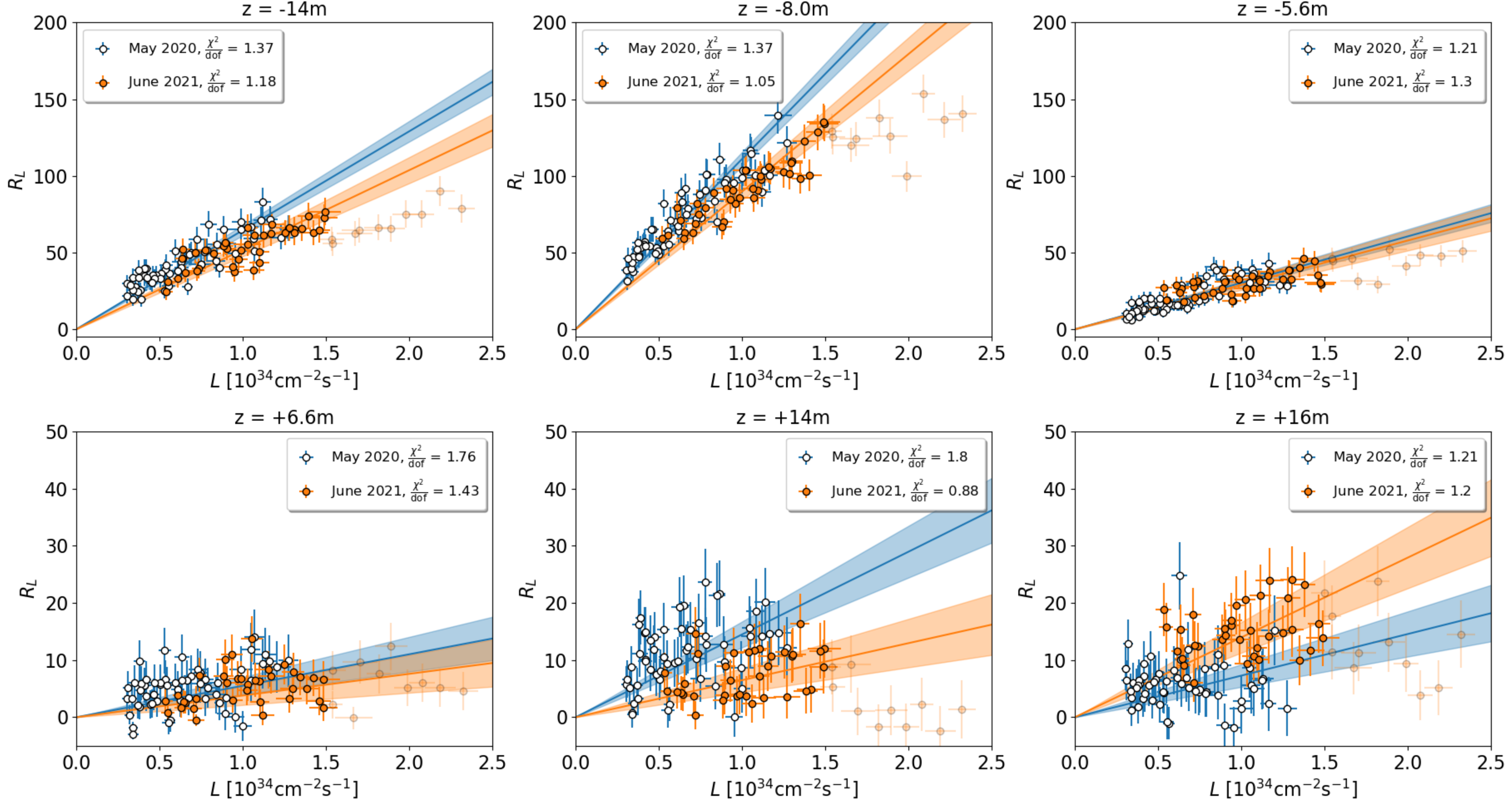}
\caption{(color online) Fits to $R_L$ versus luminosity during decay fills in each TPC. The solid blue and orange lines represent $m_L$ for Studies A and B, respectively. The shaded regions represent statistical $95\%$ confidence intervals on $m_L$ for each study. Fits to $m_L$ do not include contributions from the translucent points satisfying $L>\SI{1.5e34}{cm^{-2}s^{-1}}$ during Study B.}
\label{fig:lumi_fits}
 \end{center}
\end{figure*}

\subsection{Nuclear recoil background rates}
Eqs. (\ref{eq:heuristic}) and (\ref{eq:lumi}) provide a framework for disentangling measured backgrounds into their individual contributions. Here we use these equations to determine the observed nuclear recoil background composition and compare with simulation.

\subsubsection{Single beam background composition}
We start by applying Eq. (\ref{eq:heuristic}) to single beam LER data to measure $\alpha_\text{LER}$. To do this, we fit LER data recorded during Study A in each of the three FWD TPCs using values of $\alpha_\text{LER}$ ranging from 0.5 to 4.5 in steps of 0.1 and record the mean reduced $\chi^2$ of these fits between these three TPCs. We find a minimum reduced $\chi^2$ of 1.44 at $\alpha_\text{LER}=2.4$ with $1.8\leq\alpha_\text{LER}\leq 3.0$ representing the $95\%$ confidence interval for the value of $\alpha_\text{LER}$ that minimizes the reduced $\chi^2$ of single beam LER background fits in the FWD TPCs. In Study B, we find essentially no $\alpha_\text{LER}$ dependence on the reduced $\chi^2$ fits to single beam LER backgrounds in the FWD TPCs, but we still set $\alpha_\text{LER} = 2.4$ since $\alpha_\text{LER}$ is intended to correct for the effects of LER beam size blow up during collisions. Given that there are no appreciable HER beam size changes during collisions in both Study A and Study B, we set $\alpha_\text{HER} = 1$.
\begin{table*}[htbp]
\begin{center}
\setlength\tabcolsep{5pt}
\begin{tabular}{cccccccc}
\toprule
Study & Ring & \begin{tabular}[c]{@{}c@{}} $(\beta^*_x,\beta^*_y)$ \\ {[}\SI{}{mm},\SI{}{mm}{]} \end{tabular} & \begin{tabular}[c]{@{}c@{}}I \\ {[}mA{]} \end{tabular} & \begin{tabular}[c]{@{}c@{}} $P$ \\ {[}nPa{]} \end{tabular} &\begin{tabular}[c]{@{}c@{}} $(\sigma_y,\sigma_z)$ \\  {[}$\SI{}{\um},\SI{}{mm}${]}\end{tabular} & $n_b$ & \begin{tabular}[c]{@{}c@{}} $L$ \\ {[}$10^{34}$\SI{}{cm^{-2}s^{-1}}{]} \end{tabular} \\ \hline
\multirow{2}{*}{A} & LER & (80,1) & 510 & 30 & $(60,5.9)$  & 783 &  \multirow{2}{*}{1.1} \\
& HER  & (60,1) & 510 & 14 & $(35,6.4)$ & 783 \\  \hline
\multirow{2}{*}{B} & LER & (80,1) & 730 & 35 & $(65,5.8)$  & 1174 &  \multirow{2}{*}{2.5} \\
& HER & (60,1) &  650 & 14 & $(35,6.2)$ & 1174 \\
\bottomrule
\end{tabular}
\caption{Typical machine parameters during the luminosity background studies. All comparisons between data and simulation in this section assume these conditions.}
\label{tab:breakdown}
\end{center}
\end{table*}
\begin{table*}[htbp]
\begin{center}
\setlength\tabcolsep{5.5pt}
\begin{tabular}{clccclll}
\toprule
\begin{tabular}[c]{@{}c@{}}TPC\\$z${[}m{]} \end{tabular} & \begin{tabular}[c]{@{}c@{}} LER \\ Beam Gas \end{tabular} & \begin{tabular}[c]{@{}c@{}} HER \\ Beam Gas \end{tabular} & \begin{tabular}[c]{@{}c@{}} LER \\ Touschek \end{tabular} & \begin{tabular}[c]{@{}c@{}} HER \\ Touschek \end{tabular} & \begin{tabular}[c]{@{}c@{}} $\text{L}_\text{Study A}$ \\ {[}$^{+\text{(sys.)}}_{-\text{(sys.)}} \pm (\text{stat}.){]}$ \end{tabular}  & \begin{tabular}[c]{@{}c@{}} $\text{L}_\text{Study B}$ \\ {[}$^{+\text{(sys.)}}_{-\text{(sys.)}} \pm (\text{stat}.){]}$ \end{tabular} & $\text{Total}_\text{Study A}$\\ \hline

\rule{0pt}{2.5ex} -14 & $1.82\pm 5.2$ & 0 & N/A & $0.90\pm 0.5$ & $1.27^{+0.03}_{-0.13}\pm 0.25$ & $1.02^{+0.00}_{-0.43}\pm 0.18$ & $1.24\pm 0.22$ \\
\rule{0pt}{2.5ex} -8.0 & $9.52\pm 17$ & $39.3\pm 65.$ & N/A & $0.57\pm 0.6$ & $0.07^{+0.00}_{-0.00}\pm 0.01$ & $0.06^{+0.00}_{-0.02} \pm 0.00$ & $0.07\pm 0.01$ \\
\rule{0pt}{2.5ex} -5.6 & $22.9\pm 31$ & $16.0\pm 24.$& N/A & $0.56\pm 0.5$ & $0.14^{+0.00}_{-0.03}\pm 0.03$ & $0.13^{+0.00}_{-0.05}\pm 0.02$ & $0.14\pm 0.03$ \\
\rule{0pt}{2.5ex} +6.6 & $186.\pm 80$ & 0 & $0.98\pm 0.4$ & $4.61\pm 8.0$ & $0.14^{+0.13}_{-0.15}\pm 0.06$ & $0.10^{+0.00}_{-0.10}\pm 0.04$  & $0.44\pm 0.11$\\
\rule{0pt}{2.5ex} +14 & $471.\pm 240$ & $47.8\pm 140$ & $1.92\pm 0.5$ & N/A & $(7^{+3}_{-5}\pm 2)\times10^{-3}$ & $(3^{+0.6}_{-3} \pm 1)\times10^{-3}$ & $0.02\pm 0.00$\\
\rule{0pt}{2.5ex} +16 & $480.\pm 260$ & 0 & $1.61\pm 0.5$ & N/A & $(2^{+3}_{-2}\pm 1)\times10^{-3}$ & $(4^{+0.0}_{-4}\pm 1)\times10^{-3}$ & $0.01 \pm 0.00$ \\
\bottomrule
\end{tabular}

\caption{Data/MC ratios scaled to the conditions shown in \autoref{tab:breakdown}. Entries marked as $0$ correspond to instances where Eq. (\ref{eq:heuristic}) predicts no nuclear recoils. Entries marked as N/A indicate that no MC recoils were produced for the listed background component. Beam gas, Touschek, and total data/MC ratios were only computed for Study A. $\text{L}_\text{Study A}$ and $\text{L}_\text{Study B}$ include asymmetric systematic uncertainties accounting for potential misclassifications of single beam backgrounds.}
\label{tab:datamc}
\end{center}
\end{table*}

Using $\alpha_\text{LER} = 2.4$ and $\alpha_\text{HER}=1$, we next apply Eq. (\ref{eq:heuristic}) to 30 second averages of TPC nuclear recoil counts recorded during single beam LER and HER decay fills respectively. \autoref{fig:ler_time_fits} shows a comparison between measured recoil counts binned into $\SI{30}{s}$ intervals during the LER study period of Study A, and the corresponding LER beam-gas and LER Touschek fit predictions from Eqs. (\ref{eq:bg}) and (\ref{eq:touschek3}), respectively in the TPC located at $z_\text{BELLE} = +\SI{14}{m}$. The fit model appears to reasonably fit single beam LER backgrounds, producing a reduced $\chi^2$ of 1.3. Though not explicitly shown, we note that fits to LER and HER single beam backgrounds yield reduced $\chi^2$'s ranging between 0.8 and 1.8 in all TPCs.

\subsubsection{Luminosity background composition}

After determining $B_{0i}$, $B_{1i}$, $T_i$, and $\alpha_i$; $i=\text{LER, HER}$, we use Eq. (\ref{eq:lumi}) to determine the luminosity sensitivity parameter $m_L$ in each TPC. \autoref{fig:lumi_fits} shows linear fits to $R_L$ versus $L$ in each of the six TPCs during the decay fills of the luminosity study for both Study A and Study B. Here $R_L$ is the difference between the total number of recoils measured during each 30 second time bin during the luminosity study decay fills and the predicted number of recoils due to single beam backgrounds during this same period. We obtain reasonable fits to $R_L$ vs $L$ in all TPCs during Study A. During Study B, we observe a drop in $R_L$ at the highest luminosities (translucent points in \autoref{fig:lumi_fits}) in all TPCs. We speculate that this drop in $R_L$ above $\SI{1.5e34}{cm^{-2}s^{-1}}$ is related to the drop in effective gain at high luminosity described in \autoref{subsec:gain}. Indeed, a drop in effective gain would lower the average energy of events at high luminosity, resulting in fewer nuclear recoils above both the threshold of the chip and within our analysis selections. Furthermore, the two TPCs shown in \autoref{fig:lumi_fits} with the least significant drops in $R_L$ above $\SI{1.5e34}{cm^{-2}s^{-1}}$ during Study B are $z=\SI{-5.6}{m}$ and $z=\SI{+6.6}{m}$, which, from Eqs. (\ref{eq:cor1}) and (\ref{eq:cor2}), and \autoref{tab:calibration}, are the two TPCs that did not require gain corrections. Since it is difficult to correct for all effects related to this drop in gain, we opt to exclude the translucent points when fitting for $m_L$.
\begin{figure}
\begin{center}
\includegraphics[width=0.48\textwidth,trim={0cm 1.2cm 0cm 3cm}]{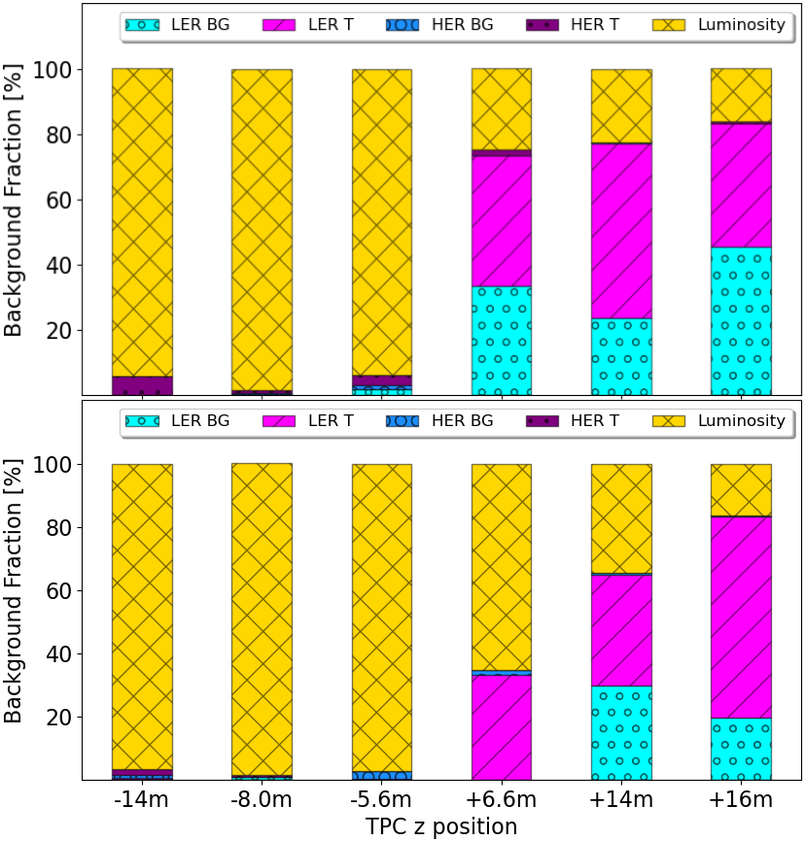}
\caption{(color online) Background compositions determined from fits to storage fills extrapolated to the machine parameters shown in \autoref{tab:breakdown} for Study A (top) and Study B (bottom).}
\label{fig:bg_breakdown}
\end{center}
\end{figure}

\subsubsection{Observed versus simulated background compositions}

\begin{figure}[htbp]
\begin{center}
\includegraphics[width=.44\textwidth,,trim={0cm .8cm 0cm 0.5cm}]{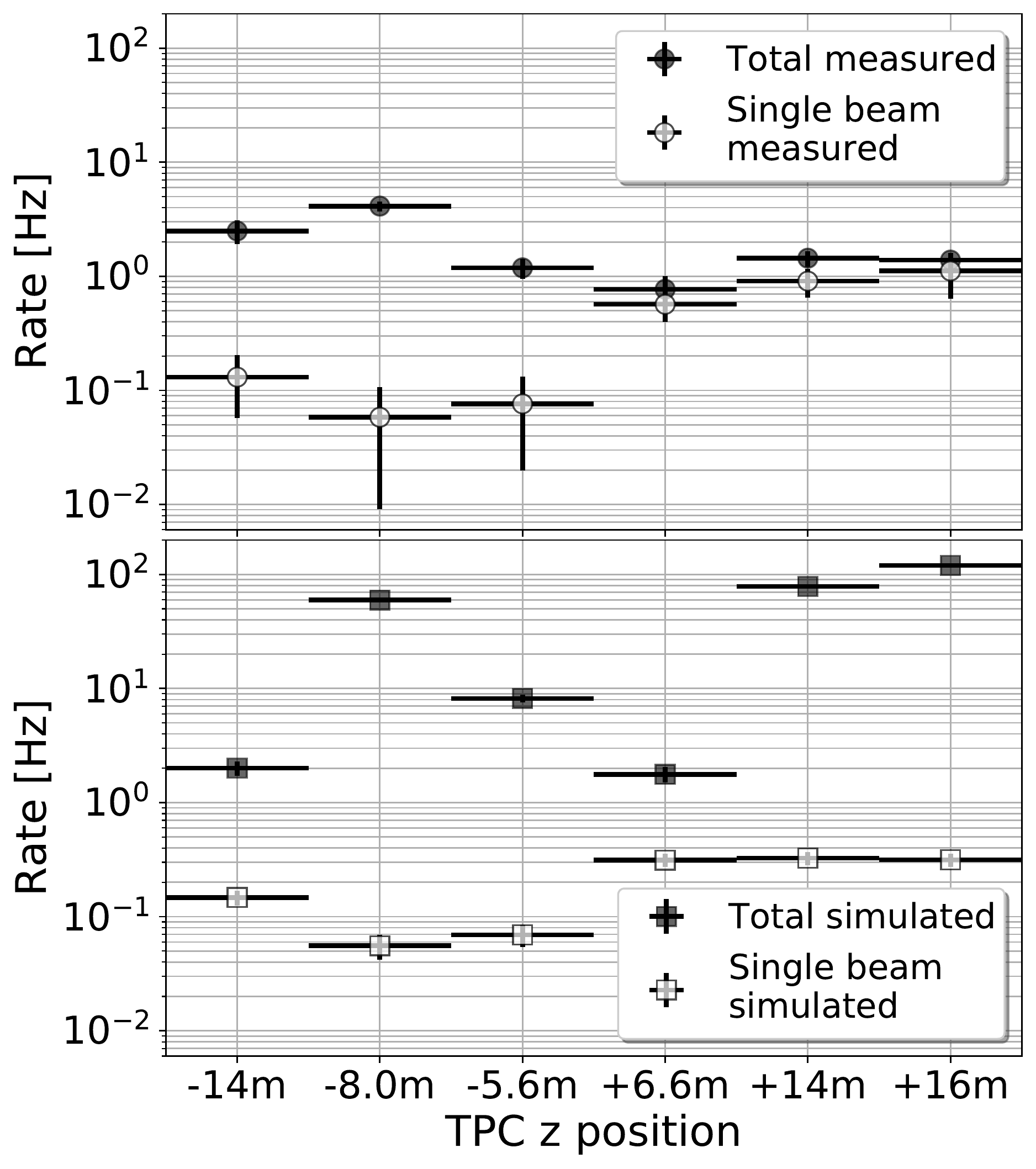}
\caption{Comparison of single beam and total (single beam + luminosity) nuclear recoil rates extrapolated to the machine conditions listed in \autoref{tab:breakdown} for observed recoils during the luminosity decay fill periods of Study A (top) and simulated recoils (bottom).}
\label{fig:rates_v_tpc}
\end{center}
\end{figure}

We next perform an extrapolation of the nuclear recoil counts for each background source observed in each TPC using machine conditions that are consistent with the decay fill periods during the luminosity study in Studies A and B. The top and bottom plots of \autoref{fig:bg_breakdown} show the extrapolated fractional background contributions at the conditions listed in \autoref{tab:breakdown} for the Study A and Study B luminosity study periods, respectively. 
Comparing the extrapolations at the conditions of Study B to the conditions of Study A, we observe that the luminosity background fraction increases in all TPCs except for the TPC at $z_\text{BELLE} = \SI{16}{m}$, where increases in LER Touschek backgrounds in Study B overshadow increases in luminosity-induced nuclear recoils at higher luminosity. Due to differences in beam optics settings that can have large effects on beam-induced background generation, directly comparing single beam backgrounds between the two background study days is not always meaningful. For a given beam optics configuration, however, we can compare observed and predicted single beam background rates. Comparing the unfilled data points in the top and bottom plots in \autoref{fig:rates_v_tpc}, we find that for the Study A beam optics configuration, observed single beam recoil rates in the BWD tunnel are consistent with prediction, while simulation under-predicts single beam recoil rates in the FWD tunnel.

\autoref{tab:datamc} shows data/MC comparisons of nuclear recoil rates for all background sources of interest. Since our simulated samples are generated assuming beam optics settings consistent with Study A, we only include this study day for single beam data/MC comparisons. We find very good agreement between data and MC Touschek backgrounds using $(\alpha_\text{LER},\alpha_\text{HER}) = (2.4,1)$ indicating that Touschek production mechanisms are modeled well in simulation. 
Furthermore, LER Touschek backgrounds appear to be the dominant single beam background source in the FWD tunnel, which is consistent with the predictions of simulation. Beam gas background measurements, on the other hand, are underpredicted by simulation. We note that simulated statistics are in general low for beam gas backgrounds leading to large uncertainties in beam gas data/MC ratios.

\begin{figure*}[htbp]
\begin{center}
\includegraphics[width=.95\textwidth]{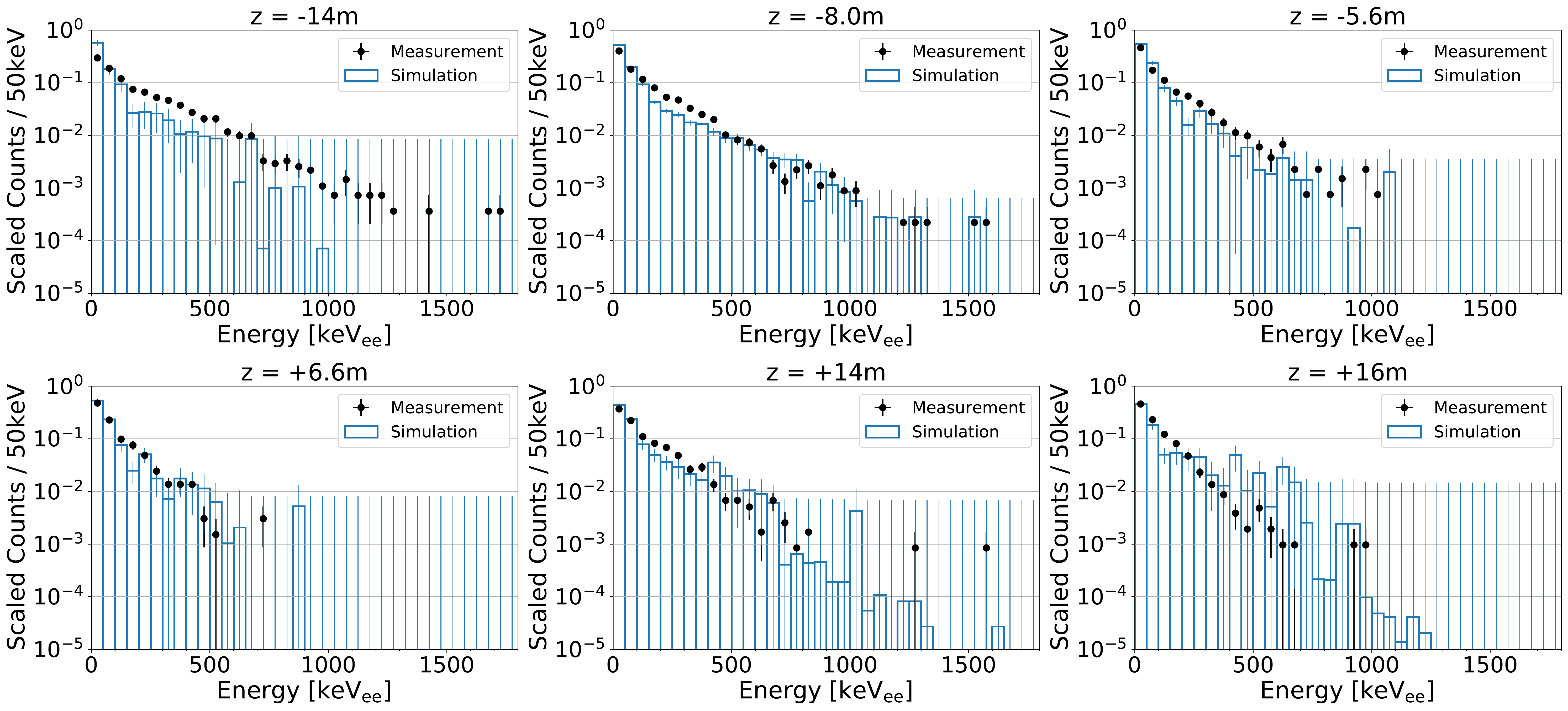}
\caption{(color online) Comparison of observed (black points) and predicted (blue bars) recoil energy spectra in each TPC. The data shown are from the beam decay luminosity study periods from Study A. The lowest energy bin only includes events above the X-ray veto threshold listed in \autoref{tab:calibration} in both measurement and simulation. Both distributions are normalized to an integral of unity. The title of each plot indicates the $z$ location of its corresponding TPC.}
\label{fig:spectra}
\end{center}
\end{figure*}

The data/MC ratios for luminosity backgrounds include an additional systematic uncertainty meant to account for potential misclassification of single beam backgrounds. These uncertainties on $m_L$ are computed assuming $1\sigma$ uncertainty contributions to $R_L$ from $B_{0i}$, $B_{1i}$, and $T_i$, that are truncated to ensure both single beam and luminosity background rates do not drop below 0. In the absence of this additional systematic uncertainty, we find agreement of luminosity data/MC ratios between Study A and Study B within $1.4\sigma$ in all TPCs. Simulation, on the other hand, predicts significantly higher luminosity background rates than measurement in all TPCs except for the TPC at $z_\text{BELLE} = -\SI{14}{m}$, as can be seen by comparing the filled data points in the top and bottom plots of \autoref{fig:rates_v_tpc}. The fact that the difference in observed luminosity recoil rates between Study A and Study B is much smaller than the difference between observed and simulated luminosity recoil rates may point to the known lack of materials such as magnet mounts, certain types of shielding, and other support structures, present in the simulated geometry description of the two accelerator tunnels, which the collaboration is currently working to improve.

\subsection{Energy spectra}
\label{subsec:spectra}
We next compare the energy spectra of measured nuclear recoils with their simulated counterparts. To perform this comparison, we first weight simulated recoil rates to provide a fair representation of the measured background composition during Study A. We apply the following steps:
\begin{enumerate}
\item Scale simulated recoils to their ``one second" (\SI{1}{s}) equivalent rates. The simulated beam time used to generate MC samples varies considerably between background sources, so we apply a scaling factor of 1/$t_\text{gen}$ where $t_\text{gen}$ is the simulated beam time, in seconds, shown in \autoref{tab:MC_sample}. Simulated luminosity recoils are further scaled down to the typical luminosity listed in \autoref{tab:breakdown}.

\item Scale simulated rates of a given background type by their corresponding data/MC factor: due to differences in data/MC ratios with respect to background type, each simulated recoil is tagged with the associated background source from which it was generated (LER/HER beam gas, LER/HER Touschek, or luminosity). Data/MC ratios are computed for recoils satisfying the X-ray veto threshold listed in \autoref{tab:calibration}, and assuming machine parameters shown in \autoref{tab:breakdown}.

\item Bin the samples by $E_\text{corrected}$.
\item Normalize to a unit integral. Since we've already compared measured and simulated recoil rates (\autoref{tab:datamc}, \autoref{fig:rates_v_tpc}), we aim to compare the shapes of the measured and simulated nuclear recoil energy spectra.
\end{enumerate}

\autoref{fig:spectra} shows the resulting histograms of measured and simulated recoil spectra. We use steps (1-3) above to assign weights to each simulated recoil.

We fill the $j$th bin of the simulated energy spectrum with the normalized sum of weights in the given bin,
\begin{align}
S_j = \frac{\sum_iw_{ji}}{\mathcal{N}},
\end{align}
where $w_{ji}$ is the weight of the $i$th event in bin $j$, and $\mathcal{N}\equiv\sum_{j}\sum_{i}w_{ji}$ is the sum of all weights in the energy spectrum. We compute the uncertainty in the $j$th bin using the following procedure:
\begin{enumerate}
\item Compute the statistical uncertainty contribution, $W_k$, from background type $k$:
\begin{align}
W_k = w_k\sqrt{N_k},
\end{align}
where $w_k$ is the weight associated with background type $k$ and $N_k$ is the number of events in the bin with background type $k$.
\item Compute the normalized uncertainty of the $j$th bin,
\begin{align}
\sigma_j = \frac{\sqrt{\sum_kW_k^2}}{\mathcal{N}}.
\end{align}
\end{enumerate}
The contents of the bins in the measured energy spectrum are unweighted. Due to measurements occurring over a substantially longer time frame than the equivalent beam time for all simulated recoil backgrounds, we include an additional ``uncertainty floor" component where we assign the quadrature sum of the weight associated with each background type, $\sqrt{\sum_kw_k^2}$, as the uncertainty of each empty bin.

With the exception of the TPC furthest away from Belle II in the BWD tunnel ($z_\text{BELLE} = -\SI{14}{m}$), we find that both the observed and simulated recoil spectra are approximately exponentially decaying with slopes in reasonable agreement, especially at energies below \SI{200}{keV}, which constitute between roughly $80\%$ and $90\%$ of recoils measured during this period in these five TPCs. The agreement between measured and predicted energy spectra in these five TPCs suggests that the material interactions that lead to neutron production are modeled well in simulation out to $\SI{16}{m}$ from Belle II in the FWD tunnel and $\SI{8.0}{m}$ in the BWD tunnel.

\subsection{Angular distributions}
\label{subsec:angular_analysis}

Comparing the distributions of measured and simulated nuclear recoil angles $\theta_\text{TPC}$ and $\phi_\text{TPC}$ provides useful insight toward our understanding of cavern neutron production points in the absence of a full kinematic reconstruction of the distribution of \textit{neutrons} incident upon a TPC. Though reconstructing incident neutron energies and angular distributions would be illuminating, doing so in a realistic detector with non-ideal performance is challenging and beyond the scope of this work. Agreement between measured and simulated recoil angles $\theta_\text{TPC}$ and $\phi_\text{TPC}$, however, should only hold if there is agreement between the angular distributions of neutrons incident upon the TPCs, so we perform such comparisons as an indirect test of our modeling of angular distributions of neutrons in the two tunnels.
\subsubsection{Axial directional performance: angular resolution}
We use the principal axis of the 3D reconstructed ionization distribution determined using an SVD (\autoref{sec:calibration}) to determine the \textit{axial} direction of a track. This ``SVD fitter" performs well for long, higher energy tracks, where the principal axis is relatively unambiguous, but for sufficiently short tracks, the reconstructed 3D track is shaped like a round bowl, making the principal axis assignment of a track ambiguous. We quantify the axial angular resolution of the TPCs by computing the angular mismeasurement (difference in angle) between the axial direction of reconstructed track using the SVD fitter and the truth MC-simulated axial direction of the recoil.
\begin{figure}[htbp]
\begin{center}
\includegraphics[width=0.425\textwidth]{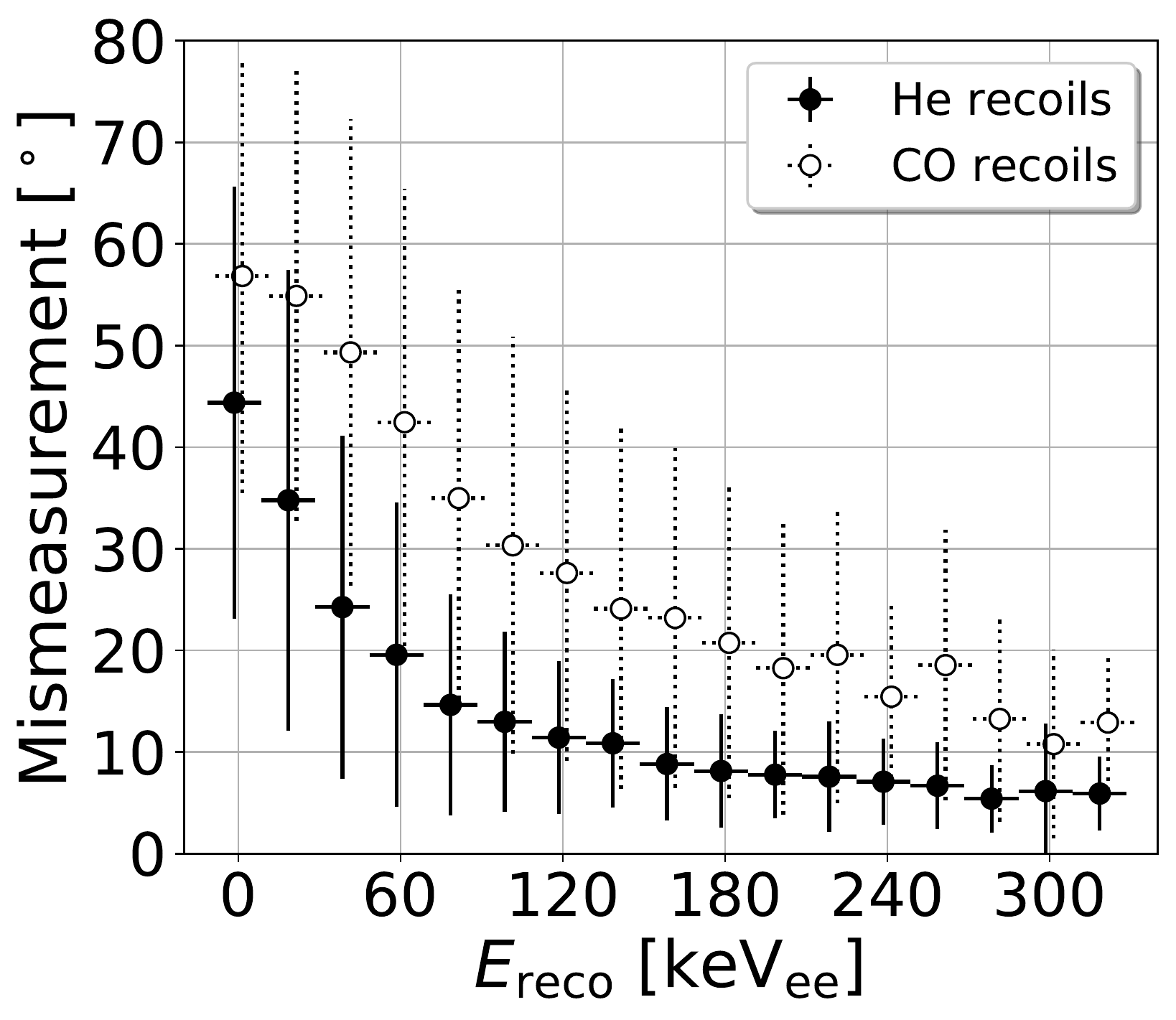}
\caption{Mean (data points) and standard deviation (error bars) of angular mismeasurement between SVD fits to the principal axis of the reconstructed simulated track and the truth MC-simulated direction of the recoil split up between $^{4}$He recoils (filled points) and $^{12}$C/$^{16}$O recoils (unfilled points). The angular resolution of $^{4}$He recoils is considerably better than $^{12}$C and $^{16}$O recoils.}
\label{fig:angres}
\end{center}
\end{figure}

\autoref{fig:angres} shows histograms of angular resolution versus $E_\text{reco}$ for truth-matched $^{4}$He and $^{12}$C/$^{16}$O recoils. We find that the angular resolution of simulated $^{4}$He recoils is considerably better than the angular resolution of $^{12}$C and $^{16}$O recoils. This finding is consistent with the fact that $^{12}$C and $^{16}$O recoils tend to produce lower energy and shorter tracks than He recoils, thus increasing the ambiguity of the identified principal axis in a measured track.

\subsubsection{Vector directionality}
\label{subsubsec:headtail}

The principal axis of a recoil track assigns the axial direction of the track, however without a \textit{vector} direction assigned to that principal axis, measurements of $\phi_\text{TPC}$ and $\theta_\text{TPC}$ are ambiguous. Here we outline a procedure for vector ``head-tail" assignment to recoil tracks and assess the performance of this procedure in assigning vector directions to tracks.

Nuclear recoils measured in a TPC are said to be ``beyond the Bragg Peak" \cite{battat}, meaning that the stopping power of an event falls sharply at the stopping end of the track, leading to an expected asymmetric distribution of charge along the principal axis of the track. Previous work with these TPCs during Phase 1 \cite{hedges,hedges2} has shown that this asymmetry is much more clear for $^{4}$He recoils than for $^{12}$C and $^{16}$O recoils, so we only attempt to assign vector directions to $^{4}$He recoil tracks. In particular, we restrict all remaining angular analyses to $^{4}$He recoils with $E_\text{reco} > \SI{40}{keV_{ee}}$ in simulation and $E_\text{corrected} > \SI{40}{keV_{ee}}$ in measurement, as simulation predicts average axial angular resolutions to be within $8^{\circ}$ for events satisfying this criteria.
\begin{figure}[htbp]
\begin{center}
\includegraphics[width=0.5\textwidth]{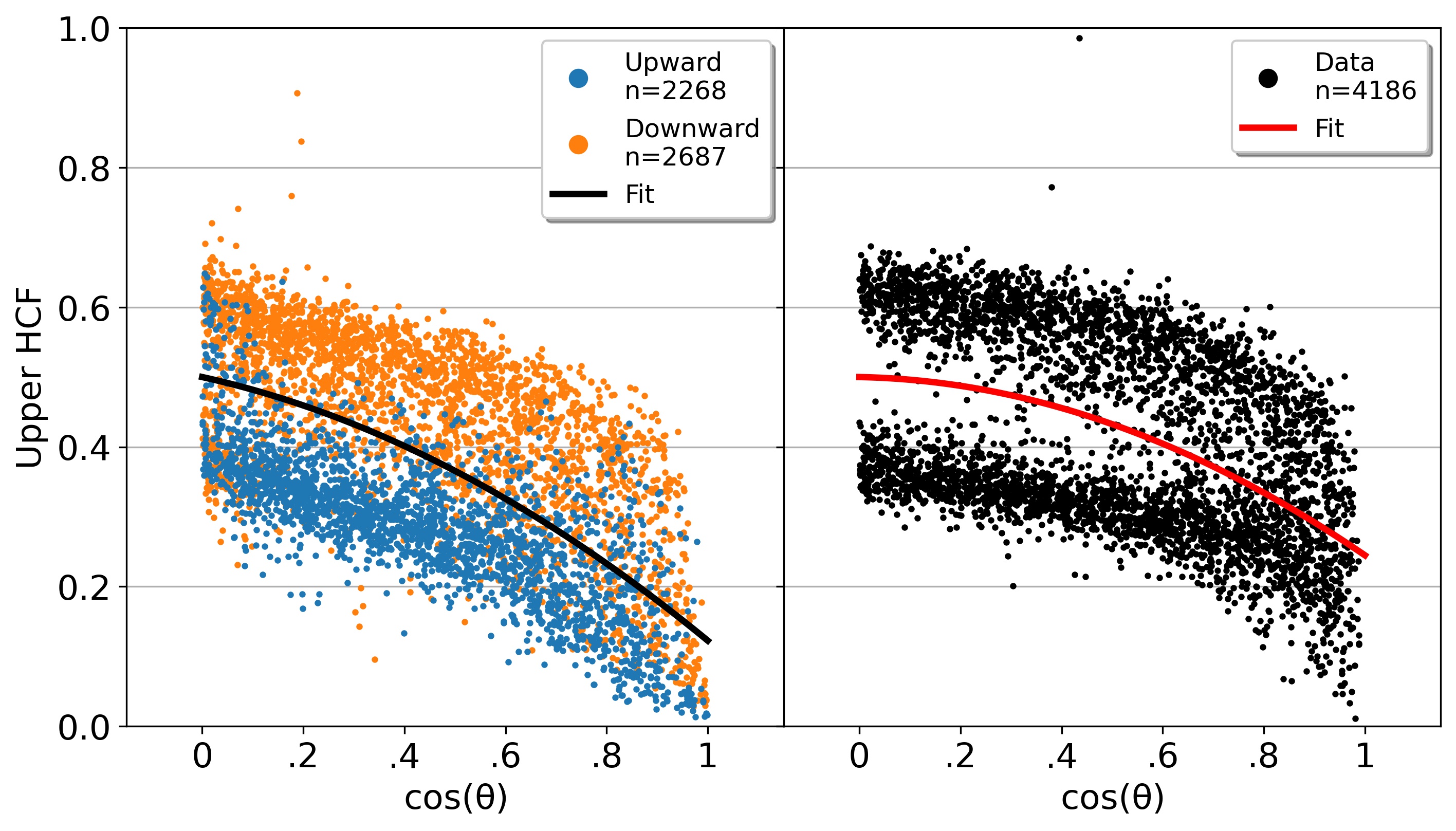}
\caption{(color online) Fractional charge of $^{4}$He recoils ($E_\text{reco} > \SI{40}{keV_{ee}}$) on the half of the track with larger $z_\text{TPC}$ (``Upper HCF") in simulation (left) and measurement (right) as a function of the inclination of the track. In both plots we find that the upper HCF tends to decrease sharply as tracks become more inclined in $z$. In simulation we observe strong separation in Upper HCF between upward pointing tracks ($\cos(\theta_\text{TPC, truth}) > 0$) and downward pointing tracks ($\cos(\theta_\text{TPC, truth}) < 0$). The black and red fit lines in the left and right plots are used as boundaries to assign vector directions to the samples of measured and simulated nuclear recoils, respectively.}
\label{fig:headtailbias}
\end{center}
\end{figure}

Charge integration effects in these TPCs are known to bias the measured charge asymmetry in a recoil away from its true ionization distribution when the recoil track is inclined with respect to the readout plane \cite{ptr}. To investigate this bias, we split each measured and simulated track in half along the midpoint of its principal axis and count up the charge on either end. We make an \textit{a priori} choice to initially define the head of the track to be the side of the track with larger average $z_\text{TPC}$ (shown in \autoref{fig:belle2_coord}). \autoref{fig:headtailbias} shows the fraction of charge on the half of each track containing the vector head (henceforth called Upper Head Charge Fraction, or Upper HCF) in all six TPCs as a function of axial track inclination\footnote{With the aforementioned initial head assignment, $0 \leq \cos(\theta_\text{TPC}) \leq 1$ so we can call $\cos(\theta_\text{TPC})$ the axial track inclination.}. We observe a steep drop in Upper HCF with increasing axial inclination in both measurement and simulation. Furthermore, two distinct Upper HCF bands arise in measured and simulated samples that are especially apparent in less inclined tracks. Simulation suggests that the band with larger (smaller) Upper HCF corresponds primarily to downward (upward) facing tracks, where downward and upward MC tracks satisfy $\cos(\theta_\text{TPC, truth}) < 0$ and $\cos(\theta_\text{TPC, truth}) > 0$, respectively. We use this separation between upward and downward facing tracks to implement our final vector head-tail assignments. The separation between these two bands appears to be stronger in measurement than in simulation, so we apply a data-driven fit boundary to implement our head-tail assignments in measurement.

The black and red fit boundaries shown for simulation and measurement, respectively in \autoref{fig:headtailbias} are determined by fitting quadratic polynomials, $p_{2,\text{ MC}}(\cos(\theta_\text{TPC}))$ and $p_{2,\text{ data}}(\cos(\theta_\text{TPC}))$ to the distribution of Upper HCF vs $\cos(\theta_\text{TPC})$ of all $^{4}$He recoils above $\SI{40}{keV_{ee}}$. In the limit of a perfectly flat track, we assume no charge integration bias in Upper HCF, so we force $p_{2,\text{ MC}}(0) = p_{2,\text{ data}}(0) = 0.5$. Using the separation between truth MC-simulated upward and downward facing tracks on either side of $p_{2,\text{ MC}}(\cos(\theta_\text{TPC}))$, we form a \textit{vector direction assignment hypothesis} that states that recoils with $\text{Upper HCF} >p_{2,\text{ MC, data}}(\cos(\theta_\text{TPC}))$ are downward facing, or equivalently, recoils with correct vector assignment have head charge fractions (HCF) less than $p_{2,\text{ MC, data}}(\cos(\theta_\text{TPC}))$. We can thus perform our final head-tail assignments by ``flipping" the vector direction (switching the vector head and tail position)
\begin{figure}[htbp]
\begin{center}
\includegraphics[width=0.5\textwidth]{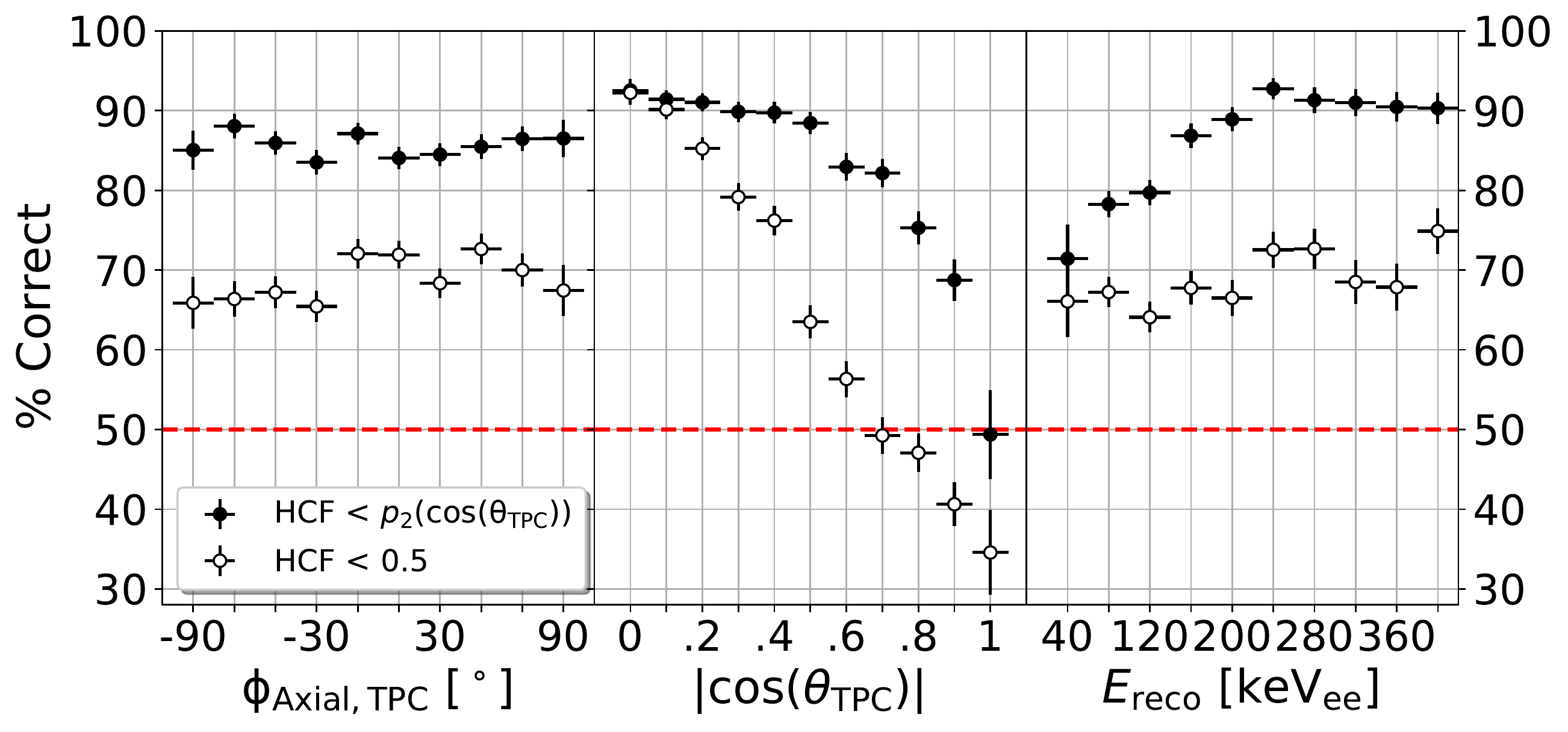}
\caption{From left to right: percentage of simulated $^{4}$He recoil tracks ($E_\text{reco} > \SI{40}{keV_{ee}}$) with assigned vector directions that match the truth MC-simulated direction of the recoil as functions of axial angle projections $\phi_{\text{Axial, TPC}}$ and $|\cos(\theta_\text{TPC})|$, and reconstructed ionization energy, $E_\text{reco}$, respectively. The filled [unfilled] points show the result using the assignment hypothesis of $\text{HCF} < p_{2,\text{ MC}}(\cos(\theta_\text{TPC}))$ [$\text{HCF} < 0.5$].}
\label{fig:headtaileff}
\end{center}
\end{figure}
of all tracks with $\text{Upper HCF} >p_{2,\text{ MC, data}}(\cos(\theta_\text{TPC}))$ to make them downward facing, thereby satisfying $\text{HCF} < p_{2,\text{ MC, data}}(\cos(\theta_\text{TPC}))$. \autoref{fig:headtaileff} shows the percentage of correctly assigned vector directions of simulated tracks verses the assigned \textit{axial} angle projections, $\phi_\text{Axial,TPC}$ and $|\cos(\theta_\text{TPC})|$, and reconstructed ionization energy $E_\text{reco}$. We find our assignment hypothesis of $\text{HCF} < p_{2,\text{ MC}}(\cos(\theta_\text{TPC}))$ (filled points in \autoref{fig:headtaileff}) leads to a significant performance improvement in vector direction assignment over the base assignment hypothesis of $\text{HCF} < 0.5$ (unfilled points in \autoref{fig:headtaileff}) which would be expected to hold in the absence of the observed charge asymmetry bias with track inclination.

After assigning vector directions to all tracks, we then compute new \textit{vector} angles $\theta_\text{TPC}'$ and $\phi_\text{TPC}'$, which are the angles of the track vector after final head-tail assignments. We shift the $\phi_\text{TPC}'$ domain to range from 0 to 360 degrees so that $180^{\circ}$ is the average direction expected for recoils caused by neutrons originating from the beam pipe. Moving forward we drop the $'$ designation of both of these recoil angles. Over our entire simulated sample of $^{4}$He recoils in all six TPCs satisfying $E_\text{reco} > \SI{40}{keV_{ee}}$ we find $\sim$ $86\%$ of recoils satisfying $\text{HCF} < p_{2,\text{ MC}}(\cos(\theta_\text{TPC}))$ have vector head assignments consistent with the true simulated direction of the recoil. If we apply an additional $90^{\circ} < \phi_\text{TPC} < 270^{\circ}$ restriction to only include recoil events with origins tending toward the beam pipe, then our percentage of correctly assigned vector directions increases to $\sim$ $91\%$.

Though we estimated our angular reconstruction performance using simulation, we argue that these estimates are reliable for measurement as well. Refs. \cite{hedges,hedges2} show consistency in axial angular resolution between measurement and simulation, finding that the angular mismeasurement between ``half-tracks", that is, tracks split in half along their principal axis, agrees to within $6^{\circ}$ between observed and simulated samples over all energies. Additionally, Ref. \cite{hedges2} shows consistency between the detected and simulated charge of recoil tracks as a function of the length along their principal axis ($\text{d}Q/\text{d}x$), indicating that the input quantities used to determine vector directional assignment are accurately simulated. Similarly, we find agreement in the shapes of $\text{d}E/\text{d}x$ distributions between measured and simulated recoils (\autoref{fig:evl}(i)) which supports that vector direction assignment is modeled accurately in simulation. Finally, as we'll see shortly, the agreement between observed and predicted angular distributions provides further evidence that our angular reconstruction performance is reliable for measured recoil tracks.
\begin{figure}[htbp]
\begin{center}
\includegraphics[width=0.5\textwidth,trim={0cm 0cm 0cm 0cm}]{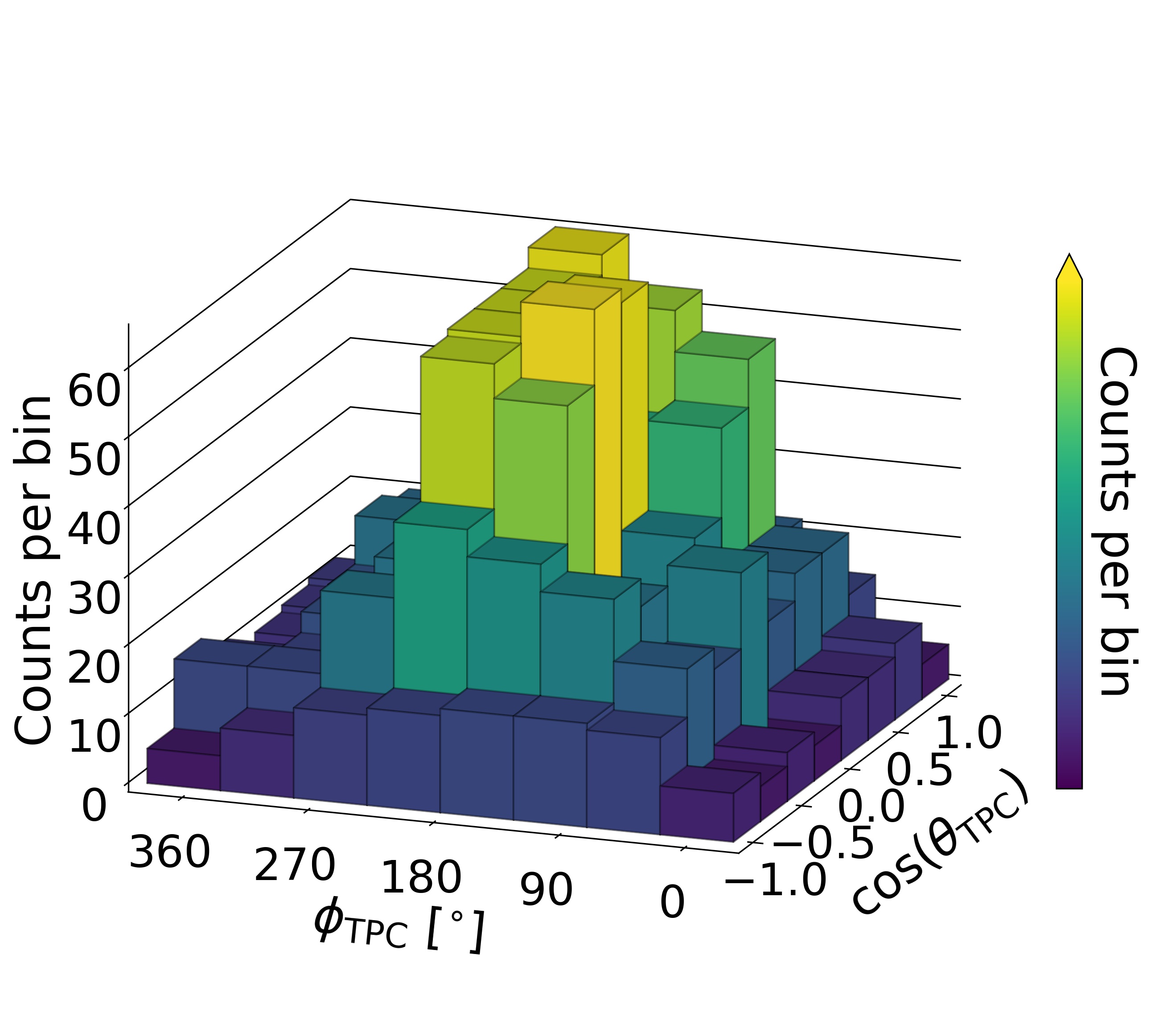}
\caption{(color online) Binned $\phi_\text{TPC}$ versus $\cos(\theta_\text{TPC})$ distribution of measured $^{4}$He recoils ($E_\text{corrected}>\SI{40}{keV_{ee}}$) after final head-tail assignment in the TPC located at $z_\text{BELLE} = -\SI{8.0}{m}$.}
\label{fig:lego}
\end{center}
\end{figure}
\begin{figure*}[htbp]
\begin{center}
\includegraphics[width=.95\textwidth, trim={2cm 1cm 1cm 2cm}]{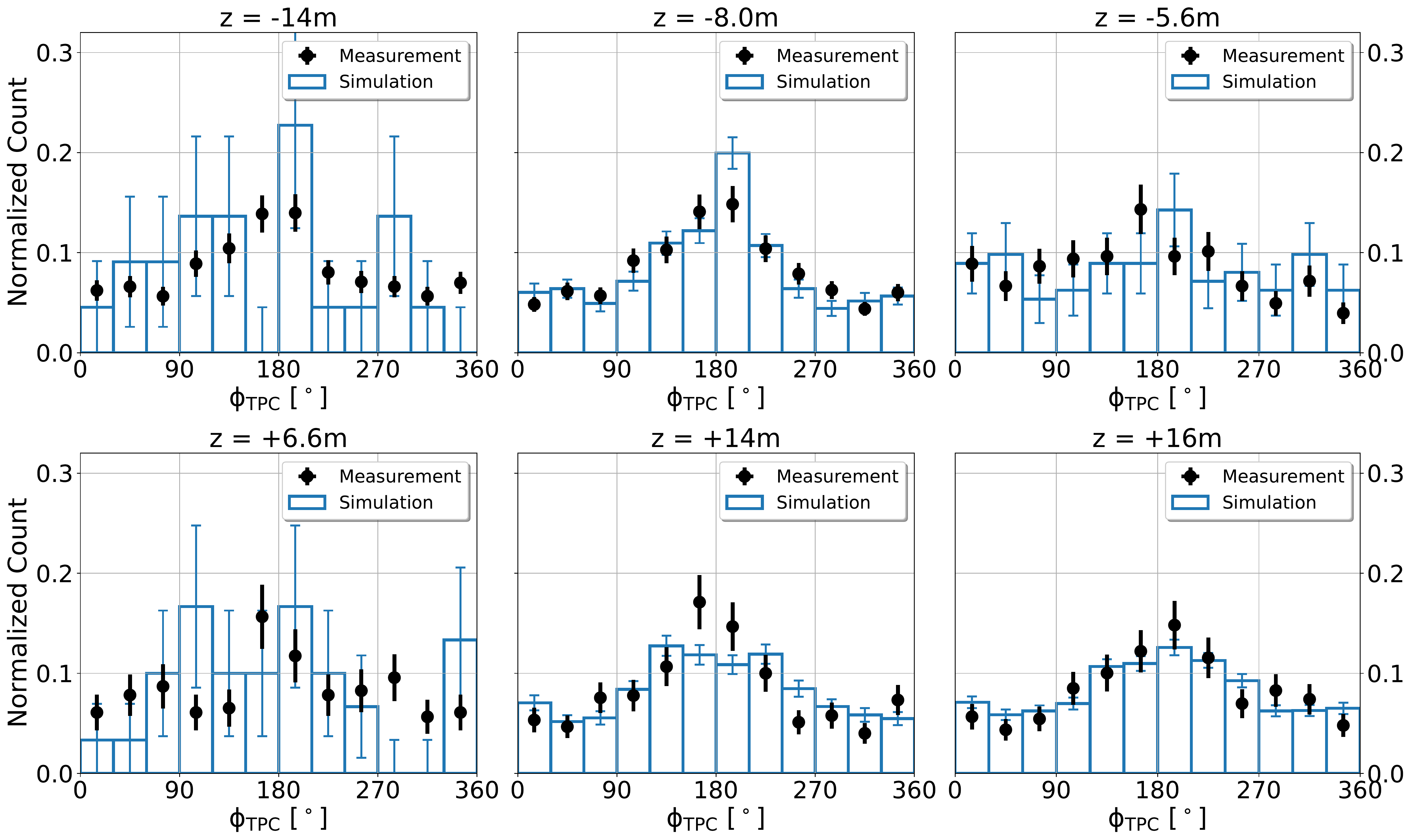}
\caption{(color online) Comparison of $\phi_\text{TPC}$ between measurement (black points) and simulated luminosity background nuclear recoils (blue bars) for events satisfying $E_\text{ionization}>\SI{40}{keV_{ee}}$ during the luminosity decay fills of Study A. Histograms are constructed after applying final directional head-tail assignments.}
\label{fig:phi_dist}
\end{center}
\end{figure*}

\begin{figure*}[htbp]
\begin{center}
\includegraphics[width=.95\textwidth, trim={2cm 1cm 1cm 2cm}]{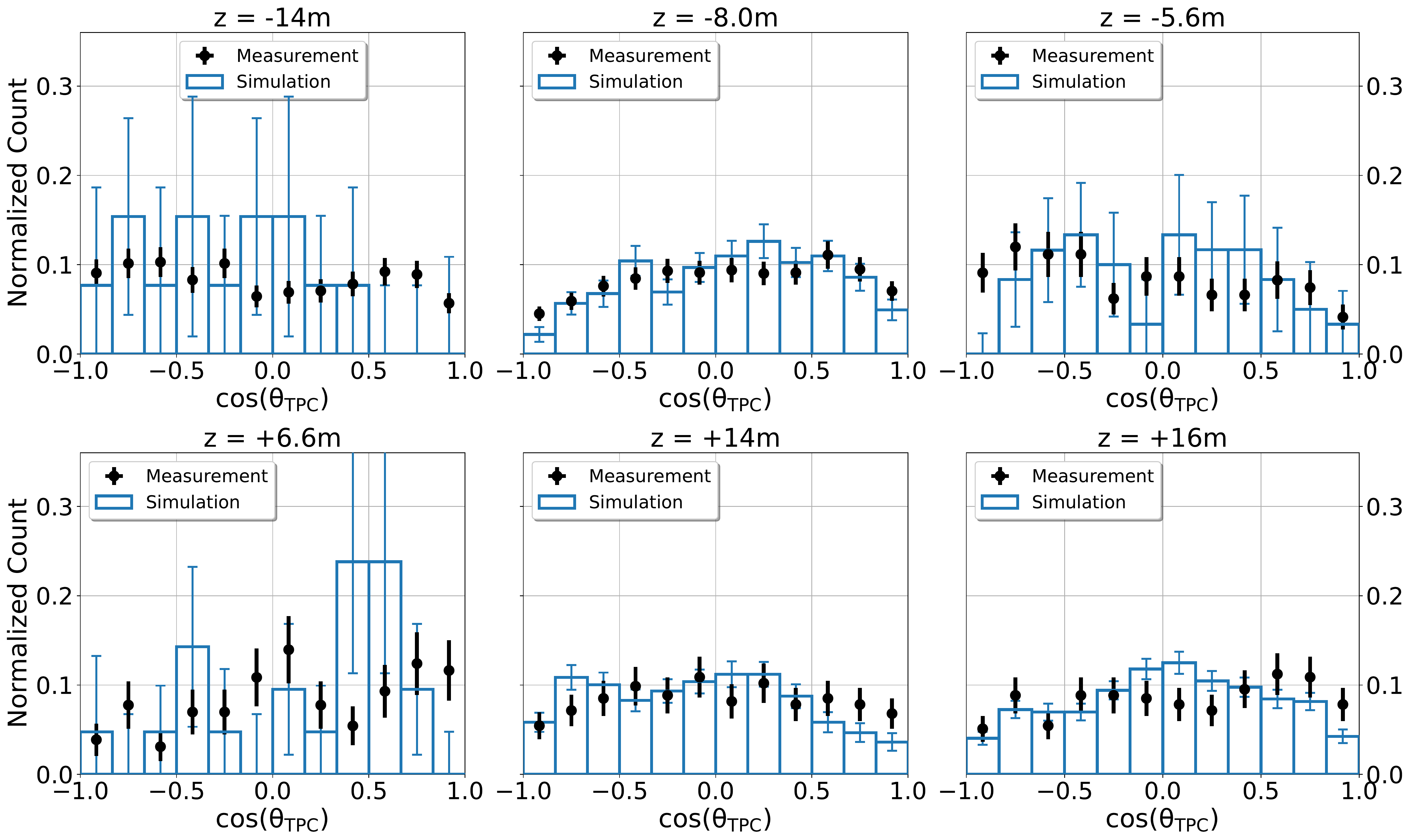}
\caption{(color online) Comparison of $\cos(\theta_\text{TPC})$ between measurement (black points) and simulated luminosity background nuclear recoils (blue bars) for events satisfying $E_\text{ionization}>\SI{40}{keV_{ee}}$ and $90^{\circ} < \phi_\text{TPC} <  270^{\circ}$ during the luminosity decay fills of Study A. Histograms are constructed after applying final directional head-tail assignments.}
\label{fig:angular_dist}
\end{center}
\end{figure*}

\subsubsection{Data versus MC comparisons of $\phi_\text{TPC}$ and $\theta_\text{TPC}$}
\autoref{fig:lego} shows a 2D histogram of $\cos(\theta_\text{TPC})$ and $\phi_\text{TPC}$ for measured $^{4}$He recoils after vector head-tail assignment satisfying $E_\text{corrected}>\SI{40}{keV_{ee}}$ from the decay fills of the Study A luminosity study period in the TPC located at $z_\text{BELLE}=-\SI{8.0}{m}$. We observe that this distribution has strong peaking along $90^{\circ} < \phi_\text{TPC} < 270^{\circ}$, consistent with the expectation that the majority of recoil events point back to the beam pipe. The shape of the $\cos(\theta_\text{TPC})$ distributions for these events coming from the beam pipe seems to also indicate that the majority of recoils point back to a cone along the beam pipe centered directly within the straight line of sight between this TPC and the beam pipe, which is qualitatively consistent with the radiative Bhabha neutron hotspot location predicted by simulation (see \autoref{fig:phase3}). With the vector directional assignment performance we've established, we can now indirectly test these claims by comparing $\phi_\text{TPC}$ and $\cos(\theta_\text{TPC})$ between measurement and simulation.

\def\figureautorefname{Figs.}\autoref{fig:phi_dist} and \ref{fig:angular_dist} show the resulting simulated $\phi_\text{TPC}$ and $\cos(\theta_\text{TPC})$ distributions plotted with measurement collected during the decay fills of the Study A luminosity study period. The histograms are normalized to an integral of unity and we report uncertainties in the recoil counts for both measurement and simulation. We also assign an uncertainty of one event (before normalization) in each empty bin. The $\cos(\theta_\text{TPC})$ distributions include an additional $90^{\circ} < \phi_\text{TPC} < 270^{\circ}$ restriction to remove events less likely to originate near the beam pipe, as neutrons originating elsewhere are not expected to be modeled well in simulation. Restricting our angular analysis to $^4$He recoils above \SI{40}{keV_{ee}} reduces the sample size of nuclear recoils substantially compared to the samples used for the energy spectra in \autoref{subsec:spectra}. To work around this limitation, we use only the luminosity background for the simulation data in \def\figureautorefname{Figs.}\autoref{fig:phi_dist} and \ref{fig:angular_dist}, as only the shape of the luminosity background angular distributions are well defined in MC after selecting directional recoils. We expect this to work well for the BWD TPCs, where luminosity background dominates, but not necessarily for the FWD TPCs.

We perform $\chi^2$ hypothesis tests \cite{gagun} comparing the normalized measured and simulated $\phi_\text{TPC}$ and $\cos(\theta_\text{TPC})$ distributions shown in \def\figureautorefname{Figs.}\autoref{fig:phi_dist} and \ref{fig:angular_dist}. In particular, we test the null hypothesis that the simulated luminosity background angular distribution explains our observed angular distributions. We test this hypothesis separately for $\phi$ and $\cos(\theta)$ and we reject the null hypothesis if the $p$-value associated with the $\chi^2$ test statistic is less than 0.05. In \autoref{tab:ang_significance}, $p_{\phi}$, and $p_{\cos(\theta)}$, are the $p$-values associated with $\chi^2$ tests for the $\phi_\text{TPC}$ and $\cos(\theta_\text{TPC})$ distributions, respectively.

\begin{table}[]
\begin{center}
\begin{tabular}{ccc}
\toprule
TPC & $p_{\phi}$ & $p_{\cos{\theta}}$ \\ \hline
$z=\SI{-14}{m}$     & 0.25 & 0.95 \\
$z=-\SI{8.0}{m}$    & 0.29 & 0.37 \\
$z=-\SI{5.6}{m}$    & 0.64 & 0.37  \\ \hline
$z=+\SI{6.6}{m}$    & 0.26 & 0.36  \\
$z=+\SI{14}{m}$     & 0.05 & 0.38  \\
$z=+\SI{16}{m}$     & 0.48 & 0.11  \\
\bottomrule
\end{tabular}
\caption{Summary of $p$-values resulting from the $\chi^2$ hypothesis tests.}
\label{tab:ang_significance}
\end{center}
\end{table}

The results of these $\chi^2$ tests suggest that at the current low level of statistics, the luminosity background alone can explain the observed distributions of recoils resulting from neutrons incident from the beam pipe. This is as expected for the BWD TPCs, where these backgrounds dominate. In the FWD TPCs, other background components are sizable, so we expect the inclusion of other backgrounds will be required to model angular distributions with larger statistics.

\subsection{Neutron flux extrapolations}
\label{subsec:extrapolation}

\begin{table*}[]
\begin{center}
\begin{tabular}{clccc}
\toprule
TPC & \begin{tabular}[c]{@{}c@{}}Luminosity \\ data/MC \end{tabular} & \begin{tabular}[c]{@{}c@{}} Scaled TPC \SI{1}{MeV} \\ Equivalent Flux\\ {[}$10^{8}$/cm$^{2}$/year{]}\end{tabular}  & \begin{tabular}[c]{@{}c@{}}Raw EKLM\\Flux \\ {[}$10^{8}$/cm$^{2}$/year{]} \end{tabular} & \begin{tabular}[c]{@{}c@{}} Scaled EKLM \SI{1}{MeV} \\ Equivalent Flux \\ {[}$10^{8}$/cm$^{2}$/year{]} \end{tabular}\\ \hline
\rule{0pt}{2.5ex} $z=-\SI{14}{m}$  & $1.27^{+0.03}_{-0.13}\pm 0.25^*$ & $1180 \pm 230$  &  \\
\rule{0pt}{2.5ex} $z=-\SI{8.0}{m}$  & $0.07^{+0.00}_{-0.00}\pm 0.01$ & $654\pm 62$     & 3.8 & 4.8                        \\
\rule{0pt}{2.5ex} $z=-\SI{5.6}{m}$   & $0.14^{+0.00}_{-0.03}\pm 0.03$  & $367\pm 69$  &                                    
\\ 
\hline
\rule{0pt}{2.5ex} $z=+\SI{6.6}{m}$ & $0.14^{+0.13}_{-0.15}\pm 0.06^*$ & $101\pm 45$    \\\
\rule{0pt}{2.5ex} $z=+\SI{14}{m}$ & $(7^{+3}_{-5}\pm 2)\times10^{-3}$ & $76 \pm 19$  & 148 &  20                   \\
\rule{0pt}{2.5ex} $z=+\SI{16}{m}$  & $(4^{+0}_{-4}\pm 1)\times10^{-3}$ & $75\pm 13$  &                                           \\
\bottomrule
\end{tabular}
\caption{Predicted luminosity neutron fluxes over one Snowmass year ($\SI{1e7}{s}$) in each TPC and in the outermost KLM endcap layers in the FWD and BWD tunnels, scaled up to SuperKEKB's target luminosity of $\SI{6.3e35}{cm^{-2}s^{-1}}$. The Raw EKLM Flux column shows expected annual neutron flux in these outermost KLM endcap layers without any measurement-informed scalings. The rightmost column shows these Raw EKLM flux estimates scaled by the corresponding highest TPC data/MC ratio in each tunnel (ratios are starred in the table).}
\label{tab:fluxes_RBB}
\end{center}
\end{table*}

We close this section off with estimates of \textit{neutron} fluxes, including those that don't produce nuclear recoils, in the tunnel regions surrounding Belle II at SuperKEKB's target luminosity of $\SI{6.3e35}{cm^{-2}s^{-1}}$. We perform these extrapolations only for luminosity backgrounds under the assumption that luminosity backgrounds are independent of beam optics settings and scale proportionately to luminosity. In the interest of extrapolating neutron fluxes in the most pessimistic scenario, we choose to perform our extrapolations assuming the larger of the two data/MC ratios computed in each TPC (\autoref{tab:datamc}).

To estimate the measured neutron flux incident on each TPC from luminosity backgrounds, we simply take the total number of luminosity-induced simulated neutrons, multiply this number by the larger of the two measured data/MC ratios for that TPC, and scale it accordingly by the relevant TPC dimensions and the simulated equivalent beam-time. We then convert this into a \SI{1}{MeV} equivalent Non-Ionizing Energy Loss (NIEL) damage-weighted flux.

We additionally scale the MC-predicted \SI{1}{MeV} equivalent NIEL damage-weighted neutron flux in the outermost BWD and FWD KLM endcap layers by the largest data/MC ratio of the TPCs in the BWD and FWD tunnels, respectively, to provide an upper limit estimate of the neutron flux reaching Belle II. \autoref{tab:fluxes_RBB} shows both the raw predicted neutron fluxes in the outermost KLM endcap layers and the scaled prediction using the largest TPC data/MC fraction in each tunnel. We note that if shielding is added between the TPCs and the KLM endcap layers, the TPC data/MC fractions should remain constant, while the simulation-predicted neutron fluxes in the outermost KLM endcaps will change, so \autoref{tab:fluxes_RBB} can still be used to predict upper limit neutron fluxes in the outermost KLM endcap layer after material description updates to simulation.

The annual upper limit neutron tolerance of the most neutron-sensitive Belle II subdetectors is $\mathcal{O}(10^{11})$ neutrons/$\text{cm}^2$/year \cite{lewis}, suggesting that all Belle II detectors are safe from luminosity-induced neutron backgrounds in the tunnel. Even so, simulation predicts that up to $95\%$ of luminosity-induced neutrons are produced within the RBB hotspot regions shown in green in \def\figureautorefname{Fig.}\autoref{fig:phase3}, so given the localized nature of neutron production during collisions, additional shielding around these RBB hotspots could be useful as a further safeguard for detector longevity.
\section{Conclusions}
\label{sec:conclusion}
We have presented the first directional measurements of fast neutron backgrounds in the tunnel region surrounding Belle II at SuperKEKB. Using an expanded simulation suite that models the SuperKEKB-Belle II geometry out to $|z_\text{BELLE}|<\SI{29}{m}$, we provided direct comparisons between measured and simulated neutrons in these tunnels surrounding Belle II. Comparing observed and simulated Touschek rates, we find agreement within a factor of 2 between data and simulation in all TPCs, indicating that Touschek production is modeled well in simulation. Beam-gas backgrounds have much larger discrepancies, most notably LER beam-gas backgrounds in the FWD tunnel where measured rates exceeded predictions of simulation by factors of up to 500. When including contributions of neutrons generated from collisions, we find agreement between total observed and predicted nuclear recoil rates within a factor of roughly $\mathcal{O}(10)$ in all TPCs except for those at $z=+\SI{14}{m}$ and $z = +\SI{16}{m}$ where simulation greatly overestimates luminosity-dependent neutron production. We note that further improvements to both the geometry and material description of collimator heads, magnets, shielding, and other components in the $|z_\text{BELLE}|<\SI{29}{m}$ tunnel region have recently been implemented in Geant4 that are not included in this work and could lead to improvements in the agreement between measured and simulated background rates. 

Comparing angular distributions of recoils, we conclude that the majority of observed and simulated neutrons originate near the beam pipe and further find reasonable agreement between measured and simulated $\cos(\theta_\text{TPC})$ distributions. Given that the simulated angular distributions only included luminosity backgrounds, and these backgrounds overwhelmingly dominate the measured recoil rates in the BWD tunnel, we suggest that these TPCs are sensitive to the predicted RBB hotspots.

When comparing the shapes of nuclear recoil energy spectra, in all except for the TPC furthest away from Belle II in the BWD tunnel, we find similar broadness between measured and simulated energy spectra. This indicates that the beam pipe and magnet material descriptions are modeled well in simulation out to at least $-\SI{8.0}{m}$ in the BWD tunnel and out to $+\SI{16}{m}$ in the FWD tunnel.

In the pursuit of investigating these neutron backgrounds, we developed a new method that removes vector directional assignment bias for highly inclined tracks, leading to the correct vector directional assignment in more than $90\%$ of simulated $^{4}$He recoils with ionization energies above $\SI{40}{keV_{ee}}$ that also point back toward the beam pipe. This same set of simulated $^{4}$He recoil events has an average angular resolution better than $8^{\circ}$. 

Finally, we have demonstrated that the recoil-imaging BEAST TPCs are capable of obtaining high purity nuclear recoil samples to reconstructed ionization energies as low as $\SI{5}{keV_{ee}}$ at effective gains of $\mathcal{O}(1000)$ (\autoref{tab:calibration}). Though we focus on operating these TPCs at low gain for the neutron background studies reported here, we note that these TPCs are capable of operating at double GEM gains approaching 50,000--a level at which they are sensitive to single electrons. Improvements in low energy directional sensitivity, angular resolution, and X-ray rejection performance are expected at higher gains, suggesting promise of using TPCs such as these in directional dark matter searches \cite{vahsenD3,cygnus,vahsennew}.

\section{Acknowledgments}
We thank Zachary Liptak for his assistance with onsite TPC operations at KEK, Majd Ghrear for providing us with TPC gas parameters, and Layne Fujioka for his contributions to visualizing 3D recoil tracks. We also thank Karsten Gadow and Caleb Miller who were both instrumental for TPC installation at KEK. Finally, we thank the anonymous referee for their insightful feedback.

We acknowledge support from the U.S. Department of Energy (DOE) via
Award Numbers DE-SC0007852, DE-SC0010504, via the U.S. Belle II Project administered by Pacific Northwest National
Laboratory (DE-AC05-76RL01830), and via U.S. Belle II Operations
administered by Pacific Northwest National Laboratory and Brookhaven
National Laboratory (DE-SC0012704).

\bibliography{neutrons_paper.bib}

\end{document}